\documentclass[a4paper,fleqn,usenatbib]{mnras}

\usepackage{newtxtext,newtxmath}
\usepackage[T1]{fontenc}
\usepackage{ae,aecompl}

\usepackage{graphicx}	% Including figure files
\usepackage{amsmath}	% Advanced maths commands
\usepackage{mathtools}
\usepackage{amssymb}	% Extra maths symbols
\usepackage{natbib}
\usepackage{hyperref}

\newcommand{\AM}{_{\rm\!AngMom}}

\newcommand{\BG}{_{_{\rm BG}}}
\newcommand{\IAD}{_{_{\rm IAD}}}
\newcommand{\MRI}{_{_{\rm MRI}}}
\newcommand{\SIS}{_{_{\rm SIS}}}
\newcommand{\SPH}{_{_{\rm SPH}}}
\newcommand{\astr}{_{_\star}}

\newcommand{\cen}{_{_{\rm CENTRAL}}}
\newcommand{\core}{_{_{\rm CORE}}}
\newcommand{\ejct}{_{_{\rm EJECT}}}
\newcommand{\env}{_{_{\rm ENVELOPE}}}
\newcommand{\esc}{_{_{\rm ESCAPE}}}
\newcommand{\fin}{_{_{\rm END}}}
\newcommand{\ijct}{_{_{\rm INJECT}}}
\newcommand{\inflow}{_{_{\rm INFLOW}}}
\newcommand{\jet}{_{_{\rm JET}}}
\newcommand{\launch}{_{_{\rm LAUNCH}}}
\newcommand{\mn}{_{_{\rm MIN}}}
\newcommand{\mx}{_{_{\rm MAX}}}
\newcommand{\neib}{_{_{\rm NEIB}}}
\newcommand{\avail}{_{_{\rm AVAIL}}}
\newcommand{\outflow}{_{_{\rm OUTFLOW}}}
\newcommand{\outt}{_{_{\rm OUT.\!1}}}
\newcommand{\rad}{_{_{\rm RAD}}}
\newcommand{\radn}{_{_{{\rm RAD}.n}}}
\newcommand{\res}{_{_{\rm RES}}}

\newcommand{\rott}{_{_{\rm ROT.\!1}}}
\newcommand{\sink}{_{_{\rm SINK}}}
\newcommand{\subO}{_{_{\rm O}}}
\newcommand{\tot}{_{_{\rm TOTAL}}}
\newcommand{\vir}{_{_{\rm VIRIAL}}}
\newcommand{\wind}{_{_{\rm WIND}}}
\newcommand{\rmsc}[1]{_{_{\rm #1}}}
\newcommand{\mrm}[1]{{{\mathrm{#1}}}}
\title[Hubble wedges in protostellar outflows]{Evolution of Hubble wedges in episodic protostellar outflows}

\author[Rohde et al.]{
P. F. Rohde,$^{1}$\thanks{E-mail: rohde@ph1.uni-koeln.de}
S. Walch,$^{1}$
D. Seifried,$^{1}$
A. P. Whitworth$^{2}$
S. D. Clarke,$^{1}$
\newauthor{ and D. A. Hubber,$^{3}$} \\
\\
$^1$ \, Universit\"at \, zu \, K\"oln, \, I. \, Physikalisches \, Institut, \,  Z\"ulpicher \, Str. \, 77, \, D-50937 \, K\"oln, \, Germany\\
$^2$ \, School \, of \, Physics \, and \, Astronomy, \, Cardiff \, University, \, Cardiff \, CF24 \, 3AA, \, UK\\
$^3$ \, Universit\"ats-Sternwarte \, M\"unchen, \, Scheinerstra{\ss}e \, 1, \, D-81679 \, M\"unchen, \, Germany
}

\date{Accepted XXX. Received YYY; in original form ZZZ}

\pubyear{2018}

\begin{document}
\label{firstpage}
\pagerange{\pageref{firstpage}--\pageref{lastpage}}
\maketitle

%%%%%
\begin{abstract}
Young low-mass protostars undergo short phases of high accretion and outburst activity leading to lumpy outflows. Recent observations have shown that the position-velocity and mass-velocity diagrams of such outflows exhibit individual bullet-like features; some of these bullets subscribe to a `Hubble Law' velocity relation, and others are manifest as `Hubble wedges'. In order to explore the origin of these features, we have developed a new episodic outflow model for the SPH code {\sc gandalf}, which mimics the accretion and ejection behaviour of FU Ori type stars. We apply this model to simulations of star formation, invoking two types of initial conditions: spherically symmetric cores in solid-body rotation with $\rho\propto r^{-2}$, and spherically symmetric turbulent cores with density proportional to the density of a Bonnor-Ebert sphere. For a wide range of model parameters, we find that episodic outflows lead to self-regulation of the ejected mass and momentum, and we achieve acceptable results, even with relatively low resolution. Using this model, we find that recently ejected outflow bullets produce a `Hubble wedge' in the position-velocity relation. However, once such a bullet hits the leading shock front, it decelerates and aligns with older bullets to form a `Hubble-law'. Bullets can be identified as bumps in the mass-velocity relation, which can be fit with a power-law, $dM/d\upsilon_{_{\rm RAD}}\propto\upsilon_{_{\rm RAD}}^{-1.5}$.
\end{abstract}
%%%%%

%%%%%
%%%%%

%%%%%
\section{Introduction}\label{chap:Introduction}
%%%%%

In the {\it core accretion} scenario for star formation, collapsing dense cores form protostars, and these protostars are usually surrounded by accretion discs \citep[e.g.][and references therein]{Shu87a, Tan14}. However, the observed luminosities of such protostars, and therefore by implication also their accretion rates, are typically much lower than one would expect, given their masses and formation time scales, and assuming steady accretion; this mismatch is called `the luminosity problem' \citep{Kenyon90,Kenyon94, Kenyon95}. The luminosity problem can be mitigated if the accretion from the disc onto the protostar is episodic, with a large fraction of the accretion occuring in short intense events called outbursts \citep{Offner11, Dunham12, Cesaroni18, Hsieh18, Ibryamov18,Kuffmeier18}. Such outbursts are observed in FU Orionis type stars (FUors), which undergo a rapid increase in accretion rate, from $ \sim 10 ^{-7}\,{\rm M}_\odot\,{\rm yr}^{-1}$ to $\sim10 ^{-4}\,{\rm M_\odot\,yr}^{-1}$, for a period of order 10 years \citep{Herbi66, Hartmann85, Hartmann89, Audard14, Safron15, Feher17}. The cause of this behaviour is still debated, but one possibility is the interplay of gravitational and magnetorotational instabilities \citep[MRI; ][]{Balbus91} in the disc \citep[e.g.][]{Zhu09,Zhu10,Stamatellos11,Kuffmeier18}. Using a large ensemble of Smoothed Particle Hydrodynamics (SPH) simulations \cite{bLomax16b,aLomax16a} conclude that episodic accretion is necessary to reproduce the observed stellar initial mass function and the ratio of brown dwarfs to hydrogen-burning stars.

Most, possibly all, forming protostars launch fast bipolar outflows (see, e.g., the reviews of \cite{Arce07, Frank14, Bally16}, or the recent results from \cite{Samal18}). The inference is that, during the collapse, gravitational energy is converted into kinetic and magnetic energy, which then drives and collimates the outflow, either through magnetic pressure or  magneto-centrifugal forces \citep{Blandford82, Lynden-Bell03, Pudritz07, Machida08, Seifried12, Frank14, Bally16}. The outflow removes a significant fraction of the angular momentum from the star-disc system, enabling the gas in the accretion disc to reach the central protostar \citep{Konigl00, Pudritz07}. Since the outflows are accretion driven and the accretion is episodic, outflows are episodic too \citep{Reipurth89, Hartigan95, Hartmann97, Konigl00, Arce07, Hennebelle11,Kuiper15, Bally16, Choi17, Cesaroni18}.

The collimated, high-velocity jet is usually traced by shock-excited H$_2$ in the early embedded phase, and later by H$\alpha$, [SII], [NII] and OI \citep[see e.g. the review by][]{Bally16}.
These jets have velocities ranging from $\sim 10\,{\rm km\,s}^{-1}$ to $\sim 150\,{\rm km\,s}^{-1}$, and carve out narrow channels \citep{Mundt83, Bally16, Liu18}. In addition, the jets are surrounded by low-velocity, wide-angle winds, with velocities up to $\sim 30\,{\rm km\,s}^{-1}$. These winds are launched further out in the accretion disc \citep{Cabrit97, Belloche02, Lee02, Arce07, Frank14}.
The collimated jet and the low-velocity, wide-angle wind entrain secondary, low-density, molecular gas, leading to a molecular outflow, mainly traced by CO and to a lesser extent by SiO \citep{Arce07}. Lines from SO and SO$_2$ trace gas that has been entrained by the wide-angle wind \citep{Tabone17}. The outflow terminates in a bow shock, where it collides with the ambient medium (the leading shock front), thus forming a shock-compressed layer of gas. Changes in the ejection rate, caused by sudden accretion events, lead to the formation of bullets and internal working surfaces, which are shocked layers between the fast ejecta and the gas in the outflow cavity. More evolved outflows break out of their parental core or cloud and form parsec-scale outflows, traced by chains of Herbig--Haro objects. In some cases these chains extend to over $10\,{\rm pc}$, e.g. HH 131 with an extent of $17\,{\rm pc}$ \citep{Reipurth97a, Reipurth98b}. The velocities and spacings of these chains of Herbig--Haro objects allow one to constrain the episodic accretion history of the launching protostar. 
\label{chap:OutflowSturcture}

A common relation observed in protostellar outflows is a linear position-velocity (PV) relation, i.e. a `Hubble Law', in which the velocity of the outflowing gas increases linearly with distance from the source \citep{Lada96}. Recent observations show that the PV diagram also exhibits so called `Hubble Wedges' of high velocity emission, caused by the bow shocks of individual outflow bullets \citep{Bachiller90, Arce01, Tafalla04, Garcia09, Wang14}. 

In addition, spectral line observations of molecular outflows show a power-law mass-velocity (MV) relation 
\begin{eqnarray}
\frac{dM}{d\upsilon\rad}&\propto&\upsilon\rad^{-\gamma}\,,
\end{eqnarray}
as first reported by \cite{Kuiper81}. While some objects show a single power-law MV relation, others can only be fit with a broken power-law \citep{Davis98, Mao14, Plunkett15,Lada96, Stojimirovic08, Qiu11,Ridge01, Birks06, Liu17}. In these cases, the low-velocity component has a shallow slope (i.e. low $\gamma_\ell$), and the high-velocity component has a much steeper slope (i.e. $\gamma_h\gg\gamma_\ell$).

\cite{Matzner99} show analytically that, for an idealised continuous hydrodynamical outflow, $\gamma_\ell$ = 2, independent of the outflow velocity, density or temperature. However, observations and simulations show that $\gamma_\ell$ can range from $\sim 1$ to $\sim 3$ with a mode of $\gamma_\ell\simeq 1.8$ \citep{Lada96, Richer00, Arce07, Plunkett15, Liu17, Li18}. \cite{Arce01} find that episodic outflows can lead to a much steeper relation, with  $\gamma_\ell\sim 2.7$. Using $1.3\,\rm{mm}$ waveband continuum and molecular line observations, \cite{Qiu09} detect jumps in the MV relation of the outflows in the high-mass star-forming region HH 80--81, and attribute these to molecular bullets, caused by episodic outflow events.

In the last decade the challenge of simulating protostellar outflows self-consistently has been tackled by a number of authors \citep{Hennebelle11, Machida09, Seifried12, Price12, Machida13, Machida14, Bate14, Tomida14, Tomida15, Lewis17}. \citet{Machida09} and \citet{Machida14} are able to reproduce the two outflow components in high resolution MHD simulations of collapsing Bonnor--Ebert spheres. However resolving the jet launching region, $r\launch\sim{\rm R}_\odot$,  and at the same time following the outflow on parsec scales, over $>\!10^5\,{\rm yrs}$, is not presently feasible computationally. Hence, these simulations are limited either to short simulation times, or to low jet velocities. An alternative approach is to invoke an almost resolution independent sub-grid model \citep{Nakamura07, Cunningham11, Peters14, Myers14, Federrath14, Offner14, Kuiper15, Offner17, Li18}. In this study, we introduce a sub-grid model which -- for the first time -- focuses on the time-variability of the outflows by mimicking the accretion behaviour of FUor type stars.  

\label{chap:OutflowVariability}
The paper is structured as follows. In Section \ref{chap:2} we describe the computational method, introduce the sub-grid outflow model and define the simulation setups. In Section \ref{chap:Study} we compare episodic and continuous outflow feedback, and discuss the results of our resolution and parameter studies, based on simulations where the initial conditions are spherically symmetric cores in solid-body rotation with $\rho\propto r^{-2}$. In Section \ref{chap:Results}, we present the results of simulations where the initial conditions are 
a spherically symmetric turbulent core with density proportional to  the density of a Bonnor-Ebert sphere, in particular concentrating on the PV and MV relations. The results are discussed in Section \ref{chap:Discussion}, and in Section \ref{chap:Conclusion} we summarise our results. 

%%%%%
\section{Computational methods and setup}\label{chap:2}
%%%%%

%%%%%
\subsection{SPH code GANDALF}\label{chap:Gandalf}
%%%%%

We use the highly object orientated `grad-h' SPH code {\sc gandalf} \citep{Hubber18}, which is based on the SPH code {\sc seren} \citep{Hubber11}. {\sc gandalf} offers different integration schemes, and we have chosen the second-order Leapfrog KDK scheme for our simulations. We use hierarchical block time-stepping to reduce the computational cost. 

{\sc gandalf} can treat the thermodynamics of the gas in several ways, for example with simple isothermal, polytropic or barotropic equations of state. Here, we invoke the approximate algorithm due to \citet{Stamatellos07}, which uses the density, $\rho_i$, temperature, $T_i$, and gravitational potential,  $\Phi_i$, of each SPH particle $i$, to estimate a mean optical depth, which is then used to compute the local heating and cooling rates. The algorithm takes into account the opacity changes due to ice mantle melting and sublimation of dust, and the switch from dust opacity to molecular line opacity; it also captures the changes of specific heat due to dissociation and ionisation of H and He.
 
To minimise the shortcomings of SPH in capturing shocks, we use the artificial viscosity formulation of \cite{Morris97},  complemented by the time-dependent artificial viscosity switch of \cite{Cullen10}.  

%%%%%
\subsection{Episodic Accretion Model}\label{chap:episodic_acc}
%%%%%

Modelling outflow feedback is crucial for star formation simulations, since outflow cavities significantly reduce the amount of gas that can be accreted onto the forming protostar, thus lowering the star formation efficiency \citep[see e.g. the review by][]{Frank14}. Self-consistent outflow simulations suffer from the extremely high spatial and temporal resolution required to resolve the jet-launching region. To bypass this problem, we treat the collapse of the core explicitly only up to the sink creation density, $\rho\sink=10^{-10}\,{\rm g\,cm}^{-3}$; thereafter we assume that a protostar forms and insert a sink particle of radius $r\sink\sim 1\,{\rm AU}$ \citep{Bate95}. We use an improved treatment of sink particles, in which SPH particles that are flagged to be assimilated by a sink particle are not assimilated instantaneously, but instead are assimilated smoothly over a few time steps. Consequently, the sink particle's surroundings are not suddenly evacuated, and this improves the hydrodynamics in the vicinity of the sink \citep{Hubber13}. In addition, the rate of inflow onto the sink, $\left.dM\sink/dt\right|\inflow$ is more smoothly varying.

Following \cite{Stamatellos12a} we divide the mass of the sink particle, $M\sink(t)$, into the mass of the central protostar, $M\astr(t)$, and the mass of an unresolved inner accretion disc, $M\IAD(t)$,
\begin{eqnarray}
M\astr(t)&=&M\sink(t)\,-\,M\IAD(t)\,.
\end{eqnarray} 
In addition, we keep track of the angular momentum of the unresolved central protostar, $L\astr(t)$, and the IAD, $L\IAD(t)$,
\begin{eqnarray}
\boldsymbol{L}\astr(t)&=&\boldsymbol{L}\sink(t)\,-\,\boldsymbol{L}\IAD(t)\,.
\end{eqnarray}
Mass and angular momentum accreted onto the sink particle are first stored in the IAD. In order to treat the episodic accretion from the IAD onto the central protostar, and the resulting accretion luminosity, we use the sub-grid episodic accretion module developed by \cite{Stamatellos12a}, based on an analytical description of FUor type stars due to \cite{Zhu09, Zhu10}, in which episodic accretion is regulated by the interplay between gravitational instability and magneto-rotational instability (MRI).

In the absence of outflow, the rate of growth of the central protostar has just two contributions,
\begin{eqnarray}\label{Eq:accretion_rate}
\frac{dM\astr}{dt}&=&\left.\frac{dM}{dt}\right|\BG\,+\,\left.\frac{dM}{dt}\right|\MRI\,.
\end{eqnarray}
The first one is the background accretion rate, $\left.dM/dt\right|\BG =10^{-7}\,{\rm M_{\odot}\, yr}^{-1}$, allowing a small amount of mass to reach the central protostar even if the MRI is not active.

The second contribution, $\left.dM/dt\right|\MRI$, represents the enhanced accretion rate enabled by the MRI during an outburst. It can exceed $\left.dM/dt\right|\BG$ by many orders of magnitude, but only during short outbursts of MRI activity. Following \cite{Zhu10b} we assume 
\begin{eqnarray}\label{eq:MRI_1}
\left.\frac{dM}{dt}\right|\MRI&=&5\times 10^{-4}\,\mathrm{M}_{\odot}\,\mathrm{yr}^{-1}\left(\frac{\alpha\MRI}{0.1}\right)\,.
\end{eqnarray}
Here, $\alpha\MRI$ is the Shakura-Sunyayev parameter \citep{Shakura73} for the effective disc viscosity due to the MRI. Simulations and observations suggest that $0.01<\alpha\MRI<0.4$ \citep{King07}. We use $\alpha\MRI\!=\!0.1$ as the default value, and vary it between $\alpha\MRI\!=\!0.05$ and $\alpha\MRI\!=\!0.2$. \cite{Zhu09b, Zhu10b} estimate that the duration of an MRI outburst is
\begin{eqnarray}\nonumber
\Delta t\MRI(t)\!\!&\!\!=\!\!&\!\!0.25\,\mathrm{kyr}\,
\left(\!\frac{\alpha\MRI}{0.1}\!\right)^{-1}\,
\left(\!\frac{M\astr(t)}{0.2 \, \mathrm{M}_{\odot}}\!\right)^{2/3}\,
\left(\!\frac{\left.dM/dt\right|\IAD}{10^{-5}\,\mathrm{M}_{\odot}\,\mathrm{yr}^{-1}}\!\right)^{1/9}\!.\\\label{eq:t_MRI}
\end{eqnarray}
Hence, the total mass deposited on the protostar during a typical MRI accretion outburst is of order
\begin{eqnarray}\nonumber
\Delta M\MRI(t)\!&\!=\!&\!\Delta t\MRI(t)\,\left.\frac{dM}{dt}\right|\MRI\\\label{Eq:M_MRI}
 \!&\!=\!&\!0.13\,\mathrm{M}_{\odot}\left(\dfrac{M\astr(t)}{0.2\,\mathrm{M}_{\odot}}\right)^{2/3}\!\left( \dfrac{\left.dM/dt\right|\IAD}{10^{-5} \, \mathrm{M}_{\odot} \mathrm{yr}^{-1}} \right)^{1/9}\!.\hspace{0.6cm}
\end{eqnarray}
\cite{Stamatellos12a} assume that, as soon as the mass of the inner accretion disc exceeds the mass for a typical MRI event, the temperature of the inner accretion disc has become high enough for thermal ionisation to activate the MRI. Hence, an outburst is triggered if 
\begin{eqnarray}\label{Eq:M_IAD}
M\IAD(t)&>&\Delta M\MRI(t)\,.
\end{eqnarray}
In the absence of outflow, the rate at which the IAD grows is given by
\begin{eqnarray}
\frac{dM\IAD}{dt}&=&\left.\frac{dM\sink}{dt}\right|\inflow-\,\left.\frac{dM}{dt}\right|\BG\,-\,\left.\frac{dM}{dt}\right|\MRI\,.\hspace{0.6cm}
\end{eqnarray}
In addition, we impose a lower limit on the mass of the IAD, $M\IAD(t)>M\mn=0.025\,{\rm M}_\odot$, to ensure that the direction of the associated angular momentum, $\boldsymbol{L}\IAD$, does not vary too much from one timestep to the next.

%%%%%
\subsection{Outflow Feedback Model}\label{SEC:Outflow}
%%%%%

Observations and theoretical predictions suggest that a fraction $f\ejct\!\sim\! 0.1 - 0.4$ of the gas accreted by a protostar is ejected in bipolar outflows \citep[but see also the review by \cite{Bally16}]{Croswell87, shu88, Pelletier92, Calvet93,  Hartmann95}. We set $f\ejct = 0.1$ as the default value. To determine the rate of outflow at the current time $t$, following a timestep $\Delta t$, we first compute the mass available for outflow,
\begin{eqnarray}\label{EQN:DeltaM_out}
\Delta M\avail(t)\!\!&\!\!=\!\!&\!\!\Delta M\avail(t\!-\!\Delta t)\,+\,f\ejct\!\left[M\astr\!(t)\!-\!M\astr\!(t\!-\!\Delta t)\right]\!.\hspace{0.8cm}
\end{eqnarray}

If the outflow model is used in combination with the episodic accretion model (see Section \ref{chap:episodic_acc}), particles are only ejected if the MRI is active (Eq. \ref{Eq:M_IAD}). Particles that would be ejected during a quiescent phase make up only a few percent compared to the outburst and are thus ejected with the particles of the outburst to improve the hydrodynamical behaviour of the outflow. If $\Delta M\avail(t) > 4 m\SPH\, $ we inject $4N(t)$ SPH particles, where
\begin{eqnarray}\label{Eq:N_part}
N(t)&=&\texttt{floor}\left(\frac{\Delta M\avail(t)}{4m\SPH}\right)\,,
\end{eqnarray}
and reduce $\Delta M\avail(t)$ accordingly,
\begin{eqnarray}
\Delta M\avail(t)&\rightarrow&\Delta M\avail(t)\,-\,4Nm\SPH\,.
\end{eqnarray} 
In Eq. (\ref{Eq:N_part}), the \texttt{floor}() operation returns the next lower natural number, and $m\SPH$ is the mass of a single SPH particle.

The SPH particles driving the outflow are injected in a cone around the angular momentum axis of the IAD, which is defined by the unit vector $\hat{\rm e}\IAD(t) = \boldsymbol{L}\IAD(t)/|\boldsymbol{L}\IAD(t)|$. They are injected in groups of four (labelled $n=1,2,3,4$), simultaneously and symmetrically (see below), firstly in order to ensure conservation of linear momentum, and secondly in order to cancel unwanted angular momentum (i.e. angular momentum that is not parallel or anti-parallel to $ \hat{\rm e}\IAD(t)$). In a spherical-polar coordinate system where the sink particle is at the origin and the polar axis is parallel to $\hat{\rm e}\IAD(t)$, the position of the first injected SPH particle in a group of four is given by $\boldsymbol{r}''_1 = (r,\theta,\phi)$, with radius $r\in[r\mn,r\mn+2\,{\rm AU}]$,  polar angle $\theta\in[0,\theta\wind]$, and azimuthal angle $\phi\in[0,2\pi]$. (All position vectors and velocity vectors in this coordinate system are distinguished by double primes, e.g. $\boldsymbol{r}''_1$.)

$r\mn$ is a purely numerical parameter, giving the smallest radius from which the SPH particles representing the outflow can be injected without incurring prohibitively short timesteps. It has default value $r\mn =10\,{\rm AU}$ (i.e. $\sim 10r\sink$; see Section \ref{chap:ejection_radius}) and must be distinguished from $r\launch$, which is the much smaller -- and unresolved --  radius from which the outflow is assumed to originate (see below). $\theta\wind$ is the opening angle of the wide-angle wind, with default value $\theta\wind =0.4\,{\rm rad}$ (see Section \ref{chap:theta_max}). $r$ and $\phi$ are drawn randomly from uniform distributions in their respective ranges, but the distribution of $\theta$ values is more complicated.

\begin{figure}
\hspace{-0.30cm}
\includegraphics[width=0.5\textwidth]{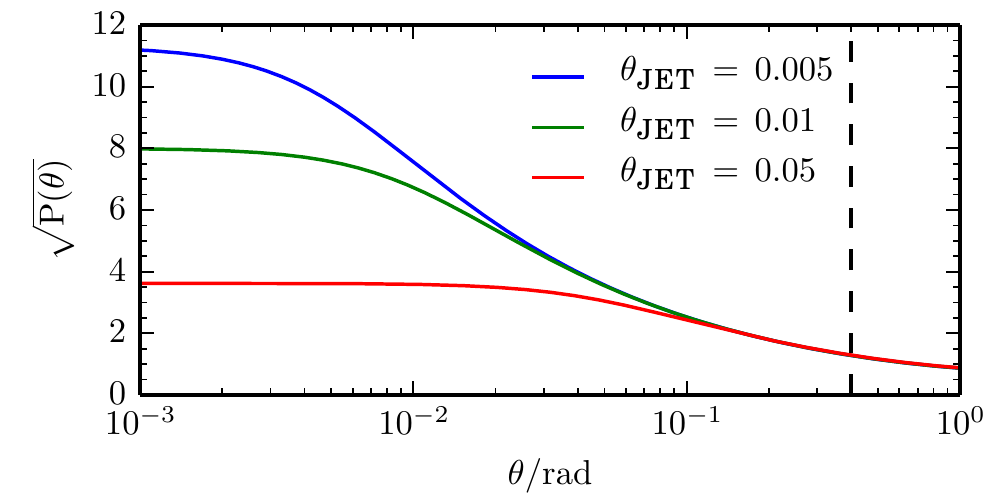}
\caption{The outflow density distribution, $\sqrt{P(\theta)}$, giving the probability that a particle is injected at polar angle $\theta$ (Eq. \ref{Eq:rho}), for three different effective opening angles $\theta\jet$. The dotted line shows the default value for the maximum polar angle, $\theta\wind =0.4$.}
\label{fig:VelProfile}  
\end{figure}

The distribution of $\theta$ determines the relative mass fractions injected in the jet- and wind-components of the outflow. Like \cite{Cunningham11, Offner14} and \cite{Kuiper15} we use the force distribution derived by \cite{Matzner99} for a hydrodynamical outflow at scales far larger than the launching region, 
\begin{eqnarray}\label{P(mu)}
P(\theta)&\propto&r^2\rho\,\upsilon^2\;\,\simeq\;\,\left[\ln\left(\frac{2}{\theta\jet}\right) \left(\sin^2(\theta)+\theta\jet^2\right)\right]^{-1}\!.\hspace{0.6cm}
\end{eqnarray} 
Here $\theta\jet$ is the jet opening angle, i.e. it determines how collimated the jet component is. We assume that the density and velocity distributions at the base of the outflow satisfy
\begin{eqnarray}\label{Eq:rho}
\rho\ijct(\theta)&\propto&P^{1/2}(\theta)\,,\\
|\upsilon\ijct(\theta)|&\propto&P^{1/4}(\theta)\,.
\end{eqnarray}
Therefore we draw $\theta$ randomly from the distribution $P^{1/2}(\theta)$ in the range $\theta\in[0,\theta\wind]$. This distribution is illustrated in Fig. \ref{fig:VelProfile} for representative values of $\theta\jet$. 

The shape of the outflow is controlled by $\theta\jet$ and $\theta\wind$. Decreasing $\theta\jet$ leads to a more collimated outflow; more particles are injected close to the outflow axis, and they are injected with higher velocities, making the jet component stronger with respect to the low-velocity wind component (see Section \ref{chap:theta_0}). Increasing $\theta\wind$ leads to a wider outflow, affecting a larger volume, with more particles being injected in the low-velocity wind component (see Section \ref{chap:theta_max}).

%%%%%
%%%%%

\begin{figure}
%\hspace{-1cm}
\includegraphics[width=0.45\textwidth]{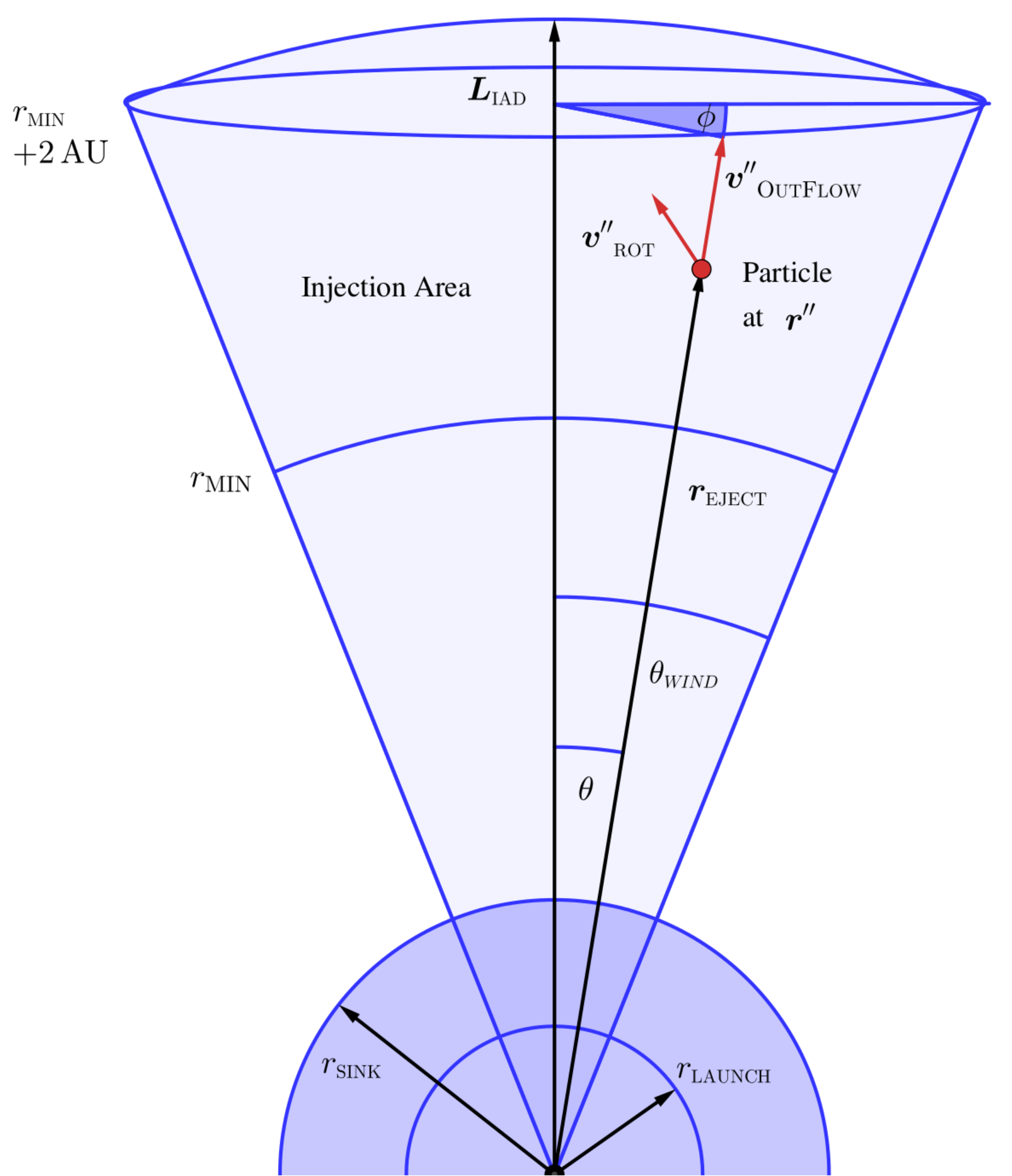}
\caption{Sketch of the outflow configuration, in the spherical polar coordinate system where the sink particle is at rest at the origin and the polar axis is parallel to $\boldsymbol{L}\IAD$. }
\label{fig:Sketch}  
\end{figure}

The outflow velocity is scaled to the Keplerian velocity at $r\launch$, i.e. $(GM\astr(t)/r\launch)^{1/2}$. Since we do not treat the evolution of the stellar radius \citep[cf.][]{Offner09}, $r\launch$ is a free parameter, with the default value $r\launch =0.07\,{\rm AU}$. We emphasise that $r\launch\ll r\sink\ll r\mn$; in other words, the SPH particles representing the outflow are injected much further out ($\geq r\mn$) than where the outflow is assumed to originate ($r\launch$). The outflow velocity, $\boldsymbol{\upsilon}''\outt$, points in the same direction as the position vector $\boldsymbol{r}''_1$ (Fig. \ref{fig:Sketch}), i.e. 
\begin{eqnarray}\label{Eq:Vout}
\boldsymbol{\upsilon}''\outt \;=\; \left(\frac{G M\astr(t)}{r\launch}\right)^{1/2}\;P^{1/4}(\theta)\;\;\frac{\boldsymbol{r}''_1}{|\boldsymbol{r}''_1|}\,,
\end{eqnarray}
where $P(\theta)$ is the force distribution (see Eq. \ref{P(mu)}). 

We add to $\boldsymbol{\upsilon}''\outt$ a rotational velocity component, $\boldsymbol{\upsilon}''\rott$, which removes angular momentum from the IAD (see Fig. \ref{fig:Sketch}). About 90\% of the associated angular momentum must be removed from the gas that reaches the protostar, so that its rotation speed matches observations and stays below the break-up limit \citep{Herbst07}. However, the physical mechanisms by which angular momentum is redistributed in a protostellar core are not fully understood and are not resolved in our simulations. The amount of angular momentum carried away by each SPH particle in the outflow is given by
\begin{eqnarray}
\ell\SPH(t)&=&f\AM\,|\boldsymbol{L}\IAD(t)|\;\frac{m\SPH}{\Delta M\MRI(t)}\,,
\end{eqnarray} 
where $f\AM = 0.9$. This ensures that the angular momentum of the protostar is a few percent of the break-up angular momentum for a $10\,{\rm R}_\odot$ protostar of the same mass. The rotational velocity component in the outflow \citep[e.g.][]{Launhardt09, Chen16, Lee17, Tabone17}  is then given by
\begin{eqnarray}
\label{Eq:v_out}
\boldsymbol{\upsilon}''\rott&=&\boldsymbol{r}''_1 \times \boldsymbol{\omega}(t)\,,
\end{eqnarray}
with 
\begin{eqnarray}
\boldsymbol{\omega}(t)&=&\frac{\ell\SPH(t)}{m\SPH\,\sin^2(\theta)\,r^2}\;\hat{\rm e}\IAD(t)\,,
\end{eqnarray}
and the additional constraint that
\begin{eqnarray}
\upsilon''\rott&\leq&\left(\frac{G M\astr(t)}{r\launch}\right)^{1/2}\,,
\end{eqnarray}
to ensure that the outflow velocity field is not dominated by rotation. If the resulting angular momentum that can be carried away by the outflow is smaller than $L\IAD$, the rest remains  in the IAD and is available for the next timestep. The total velocity of the injected SPH particle is
\begin{eqnarray}
\boldsymbol{\upsilon}''_1&=&\boldsymbol{\upsilon}''\outt + \boldsymbol{\upsilon}''\rott \,.
\end{eqnarray}

%%%%%
%%%%%

To compensate for the fact that part of the injected angular momentum is not parallel to $\hat{\bf e}\IAD(t)$, we inject a second SPH particle at position $\boldsymbol{r}''_2 = (r,\theta,\phi+\pi)$, and we compute its velocity,  analogously to the first particle, i.e.
\begin{eqnarray}
\boldsymbol{\upsilon}''_{_{\rm OUT.2}}&=&\left(\frac{G M\astr(t)}{r\launch}\right)^{1/2}\;P^{1/4}(\theta)\;\;\frac{\boldsymbol{r}''_2}{|\boldsymbol{r}''_2|}\,,\\
\boldsymbol{\upsilon}''_{_{\rm ROT.2}}&=&\boldsymbol{r}''_2 \times \boldsymbol{\omega}(t)\,,\\
\boldsymbol{\upsilon}''_2&=&\boldsymbol{\upsilon}''_{_{\rm OUT.2}}+\boldsymbol{\upsilon}''_{_{\rm ROT.2}}\,;
\end{eqnarray}
otherwise angular momentum is not conserved \citep{Hubber13}. 

The positions and velocities of particles 1 and 2 are then rotated, using a fast quaternion rotation scheme, into a frame in which the coordinate axes are parallel to those in the simulation frame, but with the origin still at the sink particle i.e. \mbox{$\boldsymbol{r}''_n \rightarrow \boldsymbol{r}'_n$} and $\boldsymbol{\upsilon}''_n \rightarrow \boldsymbol{\upsilon}'_n$, for $n=1\;{\rm and}\;2$. (All position vectors and velocity vectors in this frame are distinguished by single primes, e.g. $\boldsymbol{r}'_n$.)

To obtain a bipolar outflow, and ensure conservation of linear momentum, we inject two further SPH particles, at positions $\boldsymbol{r}'_3 = -\boldsymbol{r}'_1$ and $\boldsymbol{r}'_4 = -\boldsymbol{r}'_2$, and with velocities $\boldsymbol{\upsilon}'_3 = - \boldsymbol{\upsilon}'_1$ and $\boldsymbol{\upsilon}'_4 = - \boldsymbol{\upsilon}'_2$. The net angular and linear momenta carried away by the set of four particles are then
\begin{eqnarray}
|\Delta \boldsymbol{L}\tot(t)|\!\!&\!\!=\!\!&\!\!m\SPH\,\left|\sum_{n=1}^{n=4} \left\{\boldsymbol{r}'_n \cdot \boldsymbol{\upsilon}'_n\right\} \right|\;\,\le\;\,4\,\ell\SPH(t)\,,\hspace{0.6cm}\\
|\Delta \boldsymbol{p}\tot(t)|\!\!&\!\!=\!\!&\!\!m\SPH\,\sum_{n=1}^{n=4}\left\{\boldsymbol{\upsilon}'_n\right\}\;\,=\;\,0\,.
\end{eqnarray}

Finally, the particles are shifted into the simulation frame, with final positions and velocities
\begin{eqnarray}
\boldsymbol{r}_n&=&\boldsymbol{r}'_n + \boldsymbol{r}\sink\,,\\
\boldsymbol{\upsilon}_n&=&\boldsymbol{\upsilon}'_n + \boldsymbol{\upsilon}\sink.
\end{eqnarray} 

If more than four particles are injected ($N(t)>1$, Eq. \ref{Eq:N_part}), this process is repeated until the positions and velocities of all the new SPH particles in the outflow have been computed. Finally, if SPH particles have been injected ($N(t)>0$), the masses and angular momenta of the sink particle and its two constituent parts (the central protostar and the IAD) must be adjusted, according to
\begin{eqnarray}
M\sink(t)&\rightarrow&M\sink(t)\,-\,4N(t)m\SPH\,,\\
M\astr(t)&\rightarrow&M\astr(t)\,-\,4N(t)m\SPH\,,\\
M\IAD(t)&=&M\sink(t)\,-\,M\astr(t)\,,\\
\boldsymbol{L}\sink(t)&\rightarrow&\boldsymbol{L}\sink(t)\,-\,\Delta \boldsymbol{L}\tot(t)\,,\\
\boldsymbol{L}\astr(t)&\rightarrow&\boldsymbol{L}\astr(t)\,+\,\frac{\left(1-f\AM\right)}{f\AM}\;\Delta \boldsymbol{L}\tot(t)\,,\hspace{0.7cm}\\
\boldsymbol{L}\IAD(t)&=&\boldsymbol{L}\sink(t)\,-\,\boldsymbol{L}\astr(t)\,.
\end{eqnarray}
The model parameters are summarised in Table \ref{Table:2}.

%%%%%
\begin{table}
\caption{The model parameters, with their default values, and the ranges over which we vary them in Section \ref{chap:ParamStudy}.} 
\begin{center}\begin{tabular}{|l|l|l|}
\hline Model parameter\hspace{0.1cm} & Default value\hspace{0.1cm}           & Range \\\hline
$r\mn$ 		                   & $20\,{\rm AU}$                                     & $10\,{\rm AU\;to}\;30\,{\rm AU}$ \\ 
$r\launch$                           & $0.07\,{\rm AU}$                                  & $0.047\,{\rm AU\;to}\;0.140\,{\rm AU}$ \\ 
$\theta\wind$ 	                   & $0.4\,{\rm rad}$	                                & $0.2\,{\rm rad\;to}\,0.4\,{\rm rad}$ \\ 
$\theta\jet$ 			  & 0.01 		                                         & $0.005\;\rm{to}\;0.05$ \\ 
$f\ejct$ 	                           & 0.1 	                                                 & $0.05\;\rm{to}\;0.2$ \\  
$f\AM$ 	          & 0.9 		                                        & --- \\ 
$M\mn$		                  & $0.025\,{\rm M}_{\mathrm{\odot}}$      & ---  \\ 
\hline
\end{tabular}\end{center} 
\label{Table:2}
\end{table}
%%%%%

%%%%%
\begin{table*}
\caption{Parameters for the runs performed with the Rotating Setup (Section \ref{SEC:setup}). Reading left to right, the columns give the run number, the run ID, the feedback mechanism, the number of particles in the core (${\cal N}\core$), the sink formation density ($\rho\sink$), the wind opening angle ($\theta\wind$), the launching radius ($r\launch$), the jet opening angle ($\theta\jet$), the fraction of the gas accreted by a protostar that is ejected in bipolar outflows ($f\ejct$), the minimum radius at which the SPH particles representing the outflow are injected ($r\mn$),  and the Shakura-Sunyaev viscosity parameter for MRI viscosity ($\alpha\MRI$).} 
\begin{tabular}{|r|l|c|c|c|c|c|c|c|c|c||}
\hline $\#$ & Run & Feedback &  ${\cal N}\core$ & $\rho\sink$ & $\theta\wind$ & $r\launch$ & $\theta\jet$ & $f\ejct$ & $r\mn$ & $\alpha\MRI$  \\
 & & & $\overline{\;\;\,10^3\,\;\;}$ & $\overline{{\rm g\,cm}^{-3}}$ & $\overline{\rm radian}$ & $\overline{\rm \;\;\;\,AU\,\;\;\;}$ & $\overline{\rm radian}$ & & $\overline{\rm \;AU\;}$ & \\\hline 
 1 & {\sc Cont:100} & continuous & 100  & $10^{-10}$ & 0.4 & 0.07 & 0.01  & 0.1  & 20 & 0.1   \\ 
 2 & {\sc Cont:200} & continuous & 200  & $10^{-10}$ & 0.4 & 0.07 & 0.01  & 0.1  & 20 & 0.1   \\  
 3 & {\sc Cont:400} & continuous & 400  & $10^{-10}$ & 0.4 & 0.07 & 0.01  & 0.1  & 20 & 0.1   \\  
 4 & {\sc Episodic:25} & episodic   & 25   & $10^{-10}$ & 0.4 & 0.07 & 0.01  & 0.1  & 20 & 0.1   \\  
 5 & {\sc Episodic:50} & episodic   & 50   & $10^{-10}$ & 0.4 & 0.07 & 0.01  & 0.1  & 20 & 0.1   \\ 
 6 & {\sc Episodic:100} & episodic   &  100 & $10^{-10}$ & 0.4 & 0.07 & 0.01  & 0.1  & 20 & 0.1   \\   
 7 & {\sc Episodic:200} & episodic   & 200  & $10^{-10}$ & 0.4 & 0.07 & 0.01  & 0.1  & 20 & 0.1   \\ 
 8 & {\sc Episodic:400} & episodic   & 400  & $10^{-10}$ & 0.4 & 0.07 & 0.01  & 0.1  & 20 & 0.1   \\   
 9 & {\sc RhoSink:11} & episodic   & 100  & $10^{-11}$ & 0.4 & 0.07 & 0.01  & 0.1  & 20 & 0.1   \\ 
 10 & {\sc RhoSink:12} & episodic   & 100  & $10^{-12}$ & 0.4 & 0.07 & 0.01  & 0.1  & 20 & 0.1  \\ 
 11 & {\sc ThetaWind:0.6} & episodic   & 100  & $10^{-10}$ & 0.6 & 0.07 & 0.01  & 0.1  & 20 & 0.1  \\ 
 12 & {\sc ThetaWind:0.2} & episodic   & 100  & $10^{-10}$ & 0.2 & 0.07 & 0.01  & 0.1  & 20 & 0.1  \\ 
 13 & {\sc RLaunch:14} & episodic   & 100  & $10^{-10}$ & 0.4 & 0.14 & 0.01  & 0.1  & 20 & 0.1  \\ 
 14 & {\sc RLaunch:047} & episodic   & 100  & $10^{-10}$ & 0.4 & 0.047 & 0.01  & 0.1  & 20 & 0.1  \\ 
 15 & {\sc ThetaJet:0.05} & episodic   & 100  & $10^{-10}$ & 0.4 & 0.07 & 0.05  & 0.1  & 20 & 0.1  \\ 
 16 & {\sc ThetaJet:0.005} & episodic   & 100  & $10^{-10}$ & 0.4 & 0.07 & 0.005 & 0.1  & 20 & 0.1  \\ 
 17 & {\sc FEject:0.2} & episodic   & 100  & $10^{-10}$ & 0.4 & 0.07 & 0.01  & 0.2  & 20 & 0.1  \\ 
 18 & {\sc FEject:0.05} & episodic   & 100  & $10^{-10}$ & 0.4 & 0.07 & 0.01  & 0.05 & 20 & 0.1  \\ 
 19 & {\sc RMin:30} & episodic   & 100  & $10^{-10}$ & 0.4 & 0.07 & 0.01  & 0.1  & 30 & 0.1  \\ 
 20 & {\sc RMin:10} & episodic   & 100  & $10^{-10}$ & 0.4 & 0.07 & 0.01  & 0.1  & 10 & 0.1  \\ 
 21 & {\sc AlphaMRI:0.2} & episodic   & 100  & $10^{-10}$ & 0.4 & 0.07 & 0.01  & 0.1  & 20 & 0.2  \\ 
 22 & {\sc AlphaMRI:0.05} & episodic   & 100  & $10^{-10}$ & 0.4 & 0.07 & 0.01  & 0.1  & 20 & 0.05 \\ 
\hline 
\end{tabular} 
\label{Tab:RotIso}
\end{table*}
%%%%%

%%%%%
\subsection{Simulation Setup}\label{SEC:setup}
%%%%%

In all simulations, the initial conditions are spherically symmetric, and consist of a dense core, embedded in a low-density envelope. In the following, we consider two different initial density profiles.

For the simulations presented in Section \ref{chap:Study} (hereafter the {\it Rotating Setup}), we  construct the initial core from a singular isothermal sphere \citep[SIS;][]{Shu77}, i.e. $\rho\SIS(r)=c_{_{\rm S}}^2/2\pi Gr^2$, truncated at $R\core =0.015\,\rm{pc}$; the mass of the truncated SIS is $M=2c_{_{\rm S}}^2R\core/G=0.25\,\rm{M}_\odot$, and the density at its boundary is $\rho\SIS(R\core)=4.0\times 10^{-19}\,{\rm g\,cm}^{-3}$. Next, we increase the density of the truncated SIS by a factor of 4, so that its mass  is $M\core=1\,\rm{M}_\odot$, and it is no longer supported against self-gravity. Finally we set it in solid body rotation with angular frequency $\omega =1.35\times 10^ {-12} \,\rm{s}^{-1}$, so that it collapses to form an accretion disc with radius of order $\sim 150\,\rm{AU}$. 
 
For the simulations presented in Section \ref{chap:Results} (hereafter the {\it Turbulent Setup}), we construct the initial core from a Bonnor--Ebert sphere  \citep[BES;][]{Bonnor56, Ebert57}, since this may be a more realistic density profile than a SIS \citep{Whitworth96}. The BES has central density $\rho\cen =5\times 10^{-19}\,\rm{g\,cm}^{-3}$; it is truncated at $R\core =0.058\,\rm{pc}$, and the mass inside this radius is $1.35\,\rm{M}_\odot$. Next, we increase the density of the truncated BES by a factor of 2, so that its mass  is $M\core=2.70\,\rm{M}_\odot$, and it is no longer in hydrostatic equilibrium. Finally, we add a turbulent velocity field with the virial ratio
\begin{eqnarray}
\alpha\vir = \frac{2 (E\rmsc{TURB} + E\rmsc{THERM})}{E\rmsc{GRAV}} =  0.85 
\end{eqnarray}
and the power spectrum
\begin{eqnarray}
P_{\lambda} \propto \lambda^4 \, \mathrm{for} \, \lambda \in \left[\lambda\mn, \, \lambda\mx \right] , 
\end{eqnarray}
similar to \cite{Walch10}.
The scale of the largest, thus the most energetic, turbulent mode is set to $\lambda\mx = 2\,R\core$  \citep{Walch12}. The scale of the smallest mode is $\lambda\mn = 1 / 64 \, R\core$. 

The masses and radii of these cores resemble typical pre-stellar cores that will preferentially form low-mass protostars massive enough to launch outflows that have a significant impact on the core and envelope \citep{Motte01,Andre14}.

The dense cores are embedded in an envelope with low and uniform density, $\rho \env = 10^{-23}\,\rm{g\,cm}^{-3}$. The outer radius of the envelope is set to $R\env =1\,{\rm pc}$, in order to study multiple outflow bullets  propagating simultaneously through the envelope. In the Rotating Setup, the mass of the envelope  (between $R\core$ and $R\env$) is $M\env =0.62\,\rm{M}_\odot$, and hence $M\tot =1.62\,{\rm M}_\odot$. In the Turbulent Setup, the mass of the envelope is $M\env =1.20\,\rm{M}_\odot$, and hence $M\tot =3.90\,\rm{M}_\odot$. In both setups, the initial temperature is $T=10\,{\rm K}$ everywhere, and the corresponding isothermal sound speed is $c_{_{\rm S}}=0.19\,{\rm km\,s}^{-1}$. 

%%%%%
\begin{figure*}
\centering
\includegraphics[width=1.0\textwidth]{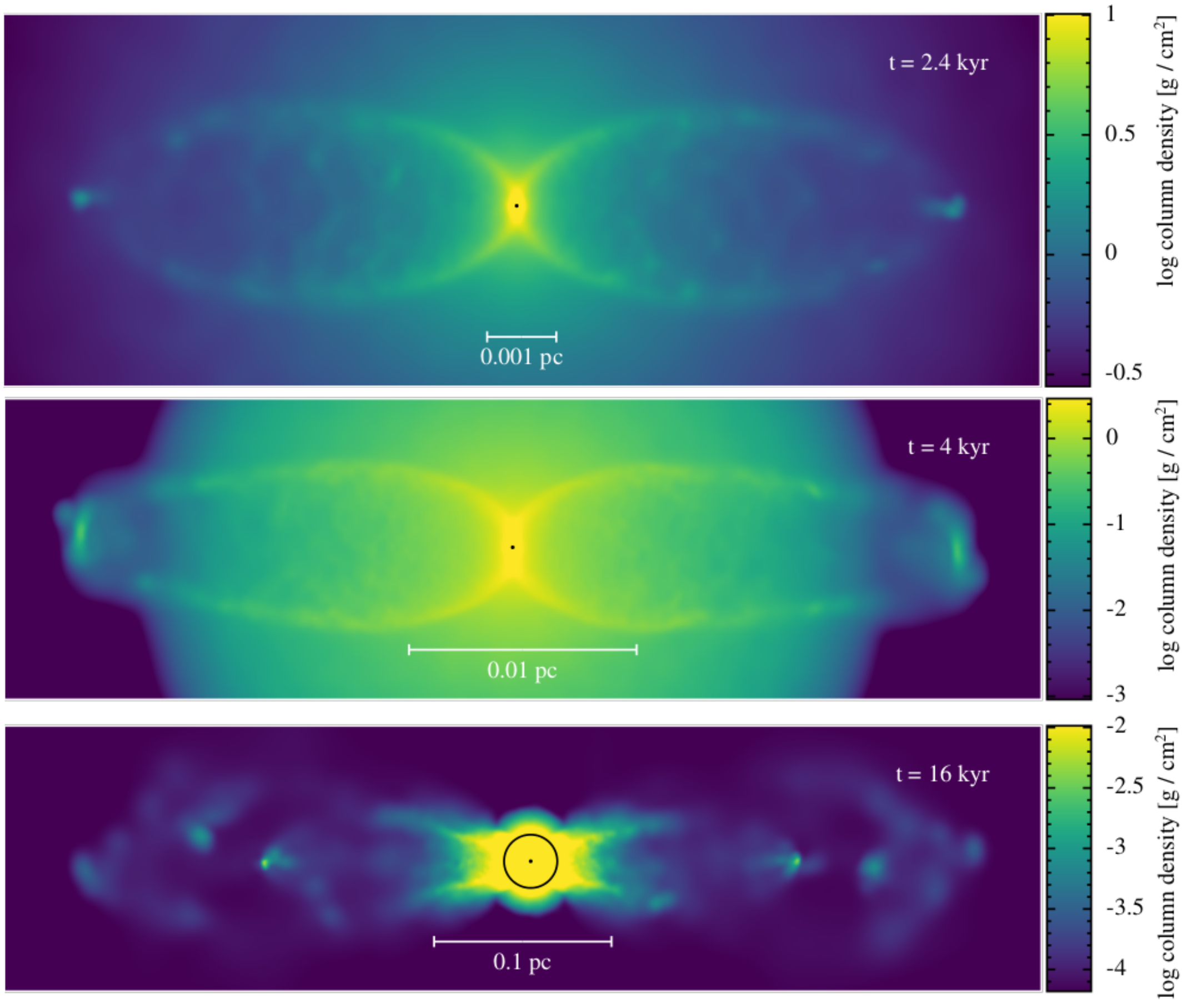}
\caption{False-colour images of the column-density in run {\sc Episodic:400} at times $t=2.4\,{\rm kyr}$, $t=4\,{\rm kyr}$ and $t=16\,{\rm kyr}$. The spatial scale increases from top to bottom, and the circle in the bottom panel shows the initial radius of the dense core ($0.015\,{\rm pc}$). The first panel shows the outflow cavity shortly after the first outflow event. In the second panel the outflow breaks out of the dense core. Multiple bullets forming a chain of Herbig--Haro-like objects are seen in the third panel.}
\label{fig:RotIso_c}
\end{figure*}
%%%%%

%%%%%
\section{The Rotating Setup}\label{chap:Study}
%%%%%

In order to illustrate the main features of core evolution, Fig. \ref{fig:RotIso_c} shows the column density from Run 8 with the Rotating Setup  (i.e. {\sc Episodic:400}, see Table \ref{Tab:RotIso}), at three different times. The first panel, at $2.4\,{\rm kyr}$, shows how the first outflow event carves out a narrow cavity. In the second panel, at $4\,{\rm kyr}$, the outflows break out of the parental dense core. In the third panel, at $16\,{\rm kyr}$, one sees multiple bullets propagating in a bipolar outflow. Since the outflow becomes longer as it evolves, the panels in Fig. \ref{fig:RotIso_c} have different scales, as shown by the scale bars.

In the following, we explore the effects of changing the numerical parameters (Section \ref{chap:Res}); of switching from episodic to continuous outflows (Section \ref{chap:Cont}); and of changing the model parameters (Section \ref{chap:ParamStudy}). The parameters for all the runs are given in Table \ref{Tab:RotIso}. For each run only one parameter is changed from its default value; the default value is the value for the fiducial case (Run 6, see Table \ref{Tab:RotIso}).

We are particularly concerned with the amount of mass and momentum escaping from the core and its envelope. An SPH particle, $n$, at radius $r_n$ (measured from the centre of mass), is deemed to have escaped if its radial velocity, $\upsilon\radn$, exceeds a notional escape velocity, i.e.
\begin{eqnarray}\label{Eq:OutfowCrit}
\upsilon\radn&>&\upsilon\esc\;\;=\;\;\left(\frac{2GM\tot}{r_n}\right)^{1/2}\,.
\end{eqnarray}
A typical value for the escape velocity at $r=0.1\,\rm{pc}$ pc is $\upsilon\esc\sim 0.37\,\rm{km\,s}^{-1}$ (where we recall that, for the Rotating Setup, $M\tot=1.62\,\rm{M}_{\odot}$).

Specifically, we evaluate, as a function of time, (i) the mass of the sink, $M\sink(t)$; (ii) the total mass carried away by {\em all} the escaping SPH particles (i.e. those that have been injected in groups of four, plus ambient SPH particles that have become entrained in the flow), i.e. 
\begin{eqnarray}
M\tot(t)&=&\sum_{n=1}^{n=N\tot(t)}\,\left\{m_n\right\}\,,
\end{eqnarray}
where $N\tot(t)$ is the total number of SPH particles to date that satisfy Eq. (\ref{Eq:OutfowCrit}); (iii) the total momentum carried away by the escaping SPH particles {\em in the outflow} (i.e. just those SPH particles that have been injected in groups of four following an accretion outburst),
\begin{eqnarray}
P\outflow(t)&=&\sum_{n=1}^{n=N\outflow(t)}\,\left\{m_n\,\upsilon\radn\right\}\,,
\end{eqnarray}
where $N\outflow(t)$ is the total number of injected SPH particles that satisfy Eq. (\ref{Eq:OutfowCrit}); and (iv) the total momentum carried away by {\em all} the escaping SPH particles, i.e.
\begin{eqnarray}
P\tot(t)&=&\sum_{n=1}^{n=N\tot(t)}\,\left\{m_n\,\upsilon\radn\right\}\,.
\end{eqnarray}

%%%%%
\begin{figure*}
\includegraphics[width=1.0\textwidth]{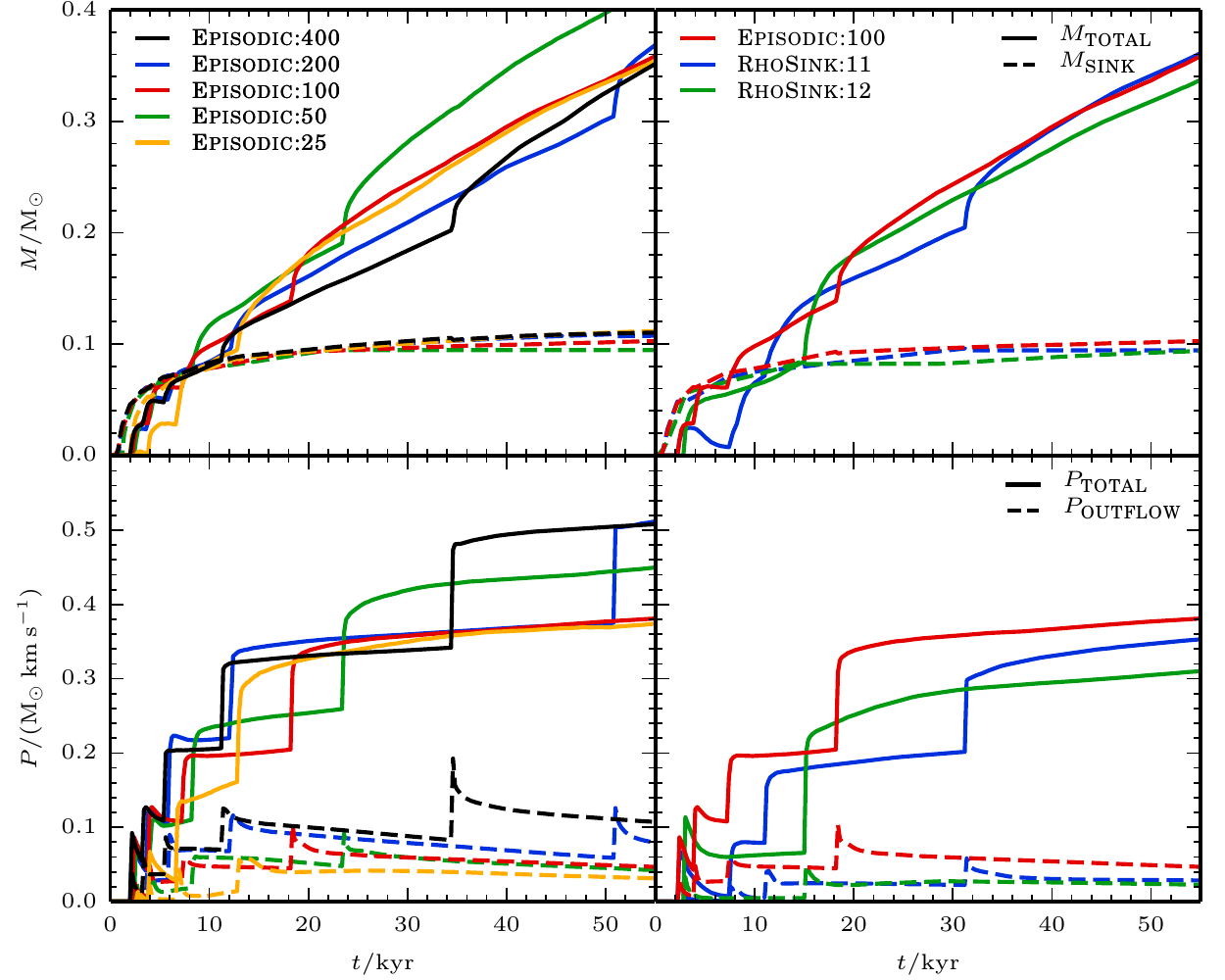}
\caption{The top panels display the evolution of the sink mass ($M\sink(t)$, dashed lines) and the total mass carried away by all the escaping SPH particles ($M\tot(t)$, solid lines). The bottom panels display the evolution of the momentum carried away by the escaping outflow SPH particles alone ($P\outflow(t)$, dashed lines), and by all the escaping SPH particles (i.e. including ambient SPH particles entrained in the flow; $P\tot(t)$, solid lines). The lefthand panels show the results obtained with different numbers of SPH particles, and hence different mass-resolution, viz. $({\cal N}\core,\,M\res) = (2.5 \times 10^4, 4 \times 10^{-3} \rm{M}_\odot)$, yellow; $(5 \times 10^4, 2 \times 10^{-3} \rm{M}_\odot)$, green; $(10^5, 10^{-3} \rm{M}_\odot)$, red; $(2 \times 10^5, 5 \times 10^{-4} \rm{M}_\odot)$, blue; $(4 \times 10^5, 2.5 \times 10^{-4} \rm{M}_\odot)$, black -- corresponding to Runs 4, 5, 6, 7 and 8, respectively. The righthand panels show the results obtained with different sink creation densities, $\rho\sink = 10^{-10}\,\rm{g\,cm}^{-3}$, red; $10^{-11}\,\rm{g\,cm}^{-3}$, blue; $10^{-12}\,\rm{g\,cm}^{-3}$, green -- corresponding to Runs 6, 9 and 10, respectively.}

\label{fig:Res}
\end{figure*} 
%%%%%

%%%%%
\subsection{Varying the numerical parameters}\label{chap:Res}
%%%%%

The Outflow Feedback Model involves two purely {\it numerical} parameters, the number of SPH particles representing the core, ${\cal N}\core$, and the density above which sink particles are created, $\rho\sink$. In this section we explore how the results change when one or other of these numerical parameters is changed from its default value (i.e. Run 6: ${\cal N}\core =10^5$, $\rho\sink =10^{-10}\,{\rm g\,cm}^{-3}$). In each of Runs 4 through 10, (see Table \ref{Tab:RotIso}), all the {\it physical} parameters are held at their default values: $\theta\wind =0.4\,{\rm rad}$, $r\launch =0.07\,{\rm AU}$, $\theta\jet =0.01\,{\rm rad}$, $f\ejct =0.1$, $r\mn =20\,{\rm AU}$ and $\alpha\MRI =0.1$.

%%%%%
\subsubsection{Mass resolution}
%%%%%

In Runs 4 through 8 (Table \ref{Tab:RotIso}), the sink creation density is held constant (at $\rho\sink =10^{-10}\,\rm{g\,cm}^{-3}$) and the number of SPH particles in the core increases from ${\cal N}\core\simeq 2.5 \times 10^4$ (Run 4) to ${\cal N}\core\simeq 4 \times 10^5$ (Run 8). Consequently the mass of an SPH particle, $m\SPH={\rm M}_\odot/{\cal N}\core$, decreases along this sequence from $m\SPH\simeq 4\times 10^{-5}\,\rm{M}_\odot$ (Run 4) to $m\SPH\simeq 2.5 \times 10^{-6}\,\rm{M}_\odot$ (Run 8). The minimum mass that can be resolved is $M\res\sim 100m\SPH$, and therefore decreases from $M\res\sim 4\times 10^{-3}\,\rm{M}_\odot$ (Run 4) to $M\res\sim 2.5 \times 10^{-4}\,\rm{M}_\odot$ (Run 8). 

Fig. \ref{fig:Res} shows the evolution of $M\sink(t)$ and $M\tot(t)$ (top left panel) and $P\outflow(t)$ and $P\tot(t)$ (bottom left panel), obtained with different ${\cal N}\core$. There are some systematic changes with increasing ${\cal N}\core$, and the results are therefore not strictly converged, even at the highest ${\cal N}\core$; in particular, the downtime between outbursts tends to be somewhat shorter with higher ${\cal N}\core$, especially at late times.

However, the overall behaviour is not strongly dependent on ${\cal N}\core$. (i) At all times the mass of the sink particle, $M\sink(t)$ varies by less than a few percent \citep[see also][]{Hubber13}. (ii) At late times ($t\ga 30\,{\rm kyr}$), the total mass ($M\tot(t)$) and momentum ($P\tot(t)$) escaping from the core and its envelope, vary with ${\cal N}\core$ (which changes by a factor of 8) by at most $20\%$. (iii) In all the simulations, the delay between successive outburst events increases with time; this is because the mass required to trigger an outburst ($\Delta M\MRI(t)$, Eq. \ref{Eq:M_MRI}) increases, and the rate at which the IAD grows ($dM\IAD/dt$) decreases, as the inflow rate from the outer parts of the core declines.

%%%%%
\subsubsection{Sink creation}
%%%%%

In Runs 6, 9 and 10 (see Table \ref{Tab:RotIso}), the number of SPH particles in the core is held constant (at ${\cal N}\core=10^5$; hence $m\SPH =10^{-5}\,{\rm M}_\odot$ and $M\res =10^{-3}\,{\rm M}_\odot$) and the sink creation density is decreased from $\rho\sink =10^{-10}\,{\rm g\,cm}^{-3}$ (Run 6) through $\rho\sink =10^{-11}\,{\rm g\,cm}^{-3}$ (Run 9) to $\rho\sink =10^{-12}\,{\rm g\,cm}^{-3}$ (Run 10). Along this sequence, the sink radius increases from $r\sink =0.9\,{\rm AU}$ (Run 6) through $r\sink =2.0\,{\rm AU}$ (Run 9) to $r\sink =4.3\,{\rm AU}$ (Run 10).

Fig. \ref{fig:Res} shows the evolution of $M\sink(t)$ and $M\tot(t)$ (top right panel), and $P\outflow(t)$ and $P\tot(t)$ (bottom right panel), obtained with different $\rho\sink$. There are some systematic changes with increasing $\rho\sink$, and the results are therefore not strictly converged, even at the highest $\rho\sink$. In particular, the downtime between outbursts is longer with lower $\rho\sink$. However, the overall behaviour is not strongly dependent on $\rho\sink$. (i) At all times the mass of the sink particle, $M\sink(t)$ varies by less than a few percent. (ii) In all the simulations, the downtime between successive outburst events increases with time. (iii) At late times ($t\ga 30\,{\rm kyr}$), as $\rho\sink$ changes by a factor of 100, the total mass escaping from the core and its envelope, $M\tot(t)$, varies by at most $10\%$; and the total momentum escaping, $P\tot(t)$, by at most $20\%$. 

%%%%%
\subsubsection{Synopsis}
%%%%%

Low ${\cal N}\core$ means not having to follow so many SPH particles (but at the price of coarser mass resolution), and low $\rho\sink$ means not having to follow the SPH particles to such high densities, and hence with such short timesteps (but at the price of excising the detailed dynamics at these high densities). On both counts this reduces computing requirements. Since our outburst model produces approximately converged results with low ${\cal N}\core$ and low $\rho\sink$ -- modulo slightly longer downtimes between outburst events -- we conclude that it can be used in larger scale simulations of star formation \citep[e.g.][]{Clarke17}, where the computing requirements with higher ${\cal N}\core$ and/or higher $\rho\sink =10^{-10}\,{\rm g\,cm}^{-3}$ might not be feasible.

%%%%%
\subsection{Comparison between episodic and continuous outflows}\label{chap:Cont}
%%%%%

Here, we compare and contrast episodic outflow feedback with continuous outflow feedback. The continuous feedback runs (Runs 1 to 3, Table \ref{Tab:RotIso}) equate the accretion rate onto the protostar ($dM\astr/dt$) to the accretion rate onto the sink particle ($dM\sink/dt$) -- rather than using the sub-grid model of \cite{Stamatellos12} and obtaining episodic accretion rate onto the protostar; we reiterate that in the episodic approach particles are only ejected during an outburst (see Section \ref{SEC:Outflow}). The continuous feedback runs use ${\cal N}\core\simeq 10^5$ (Run 1), ${\cal N}\core\simeq 2\times 10^5$ (Run 2), and ${\cal N}\core\simeq 4\times 10^5$ (Run 3); all other parameters have their default values. The results are compared with the episodic accretion Runs 6 and 7. 

%%%%%
\begin{figure}
\hspace{-0.5cm}
\centering
\includegraphics[width=0.5\textwidth]{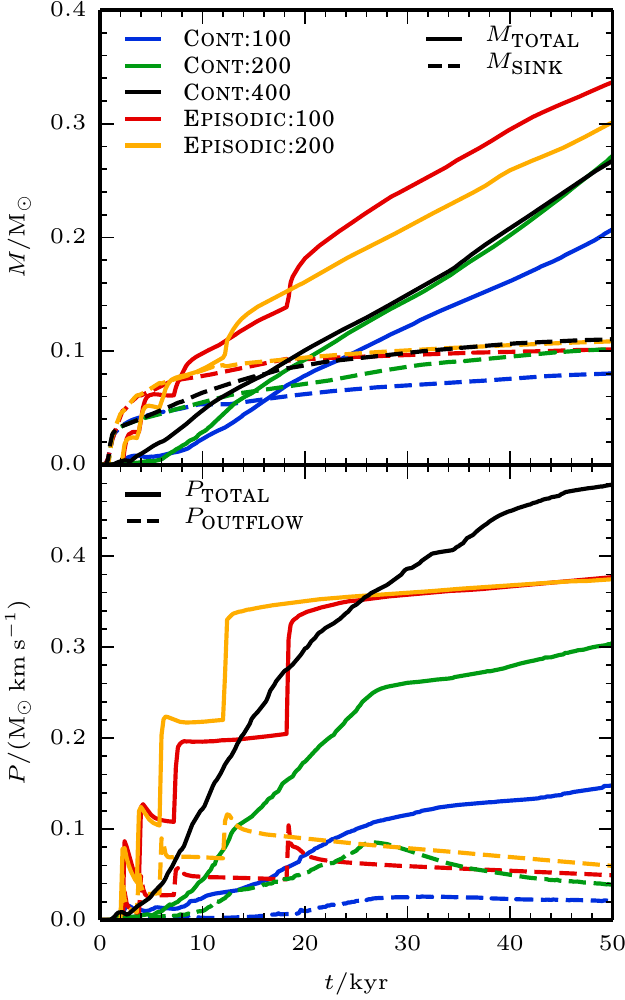}
\caption{As Fig. \ref{fig:Res}, but for Run 1 ({\sc Cont:100}; blue), Run 2 ({\sc Cont:200}; green), Run 3 ({\sc Cont:400}; black), Run 6 ({\sc Episodic:100}; red) and Run 7 ({\sc Episodic:200}; yellow), comparing and contrasting episodic and continuous outflow feedback.}
\label{fig:cont}
\end{figure}
 %%%%%
 
Fig. \ref{fig:cont} shows the evolution of $M\sink(t)$ and $M\tot(t)$ (top panel), and $P\outflow(t)$ and $P\tot(t)$ (bottom panel), obtained with continuous and episodic outflows. There is no evidence that the runs with continuous outflows have converged with increasing ${\cal N}\core$ -- particularly as regards the momentum carried away from the core (blue, green and black lines on the bottom panel of Fig. \ref{fig:cont}) -- and therefore we cannot safely draw any conclusions regarding the physics of continuous outflows. The reason for non-convergence is that the continuous outflows are not properly resolved.

To be properly resolved, the SPH particles near the axis of the outflow must have neighbour-lists which contain exclusively SPH particles that are also in the outflow (and {\it all} SPH particles in the outflow must have neighbour-lists which contain at least a significant fraction of SPH particles that are also in the outflow). In other words, the SPH particles in the outflow must interact hydrodynamically with one another. With continuous outflow, the rate at which outflow SPH particles are launched may be so low that this requirement is not met; successively ejected outflow SPH particles may be too far apart. The SPH particles in the outflow will only interact hydrodynamically with one another if they have very low-mass, and therefore are very numerous, specifically
\begin{eqnarray}\label{EQN:mSP}
m\SPH\!\!&\!\!<\!\!&\!\!\frac{\dot{M}\astr\,f\ejct\,(4\pi/3)\,r\mn}{4\,\bar{\cal N}\neib\,\pi\theta\jet^2\,\upsilon\launch}\\
\!\!&\!\!<\!\!&\!\!7\times 10^{-6}\,{\rm M}_\odot\left(\frac{\dot{M}\astr}{10^{-6}\,{\rm M}_\odot\,{\rm yr}^{-1}}\right)\left(\frac{\upsilon\launch}{100\,{\rm km\,s}^{-1}}\right)^{\!-1}\!.\hspace{0.9cm}
\end{eqnarray}
To obtain the second expression, we have substituted the default values for $f\ejct=0.1$, $r\mn=20\,{\rm AU}$, and $\theta\jet =0.01\,{\rm rad}$. The factor 4 in the denominator derives from the fact that we launch outflow SPH particles in groups of four. $\bar{\cal N}\neib$ is the mean number of neighbours, and we have substituted $\bar{\cal N}\neib =50$. In the runs with continuous outflow, $\dot{M}\astr$ is usually much smaller than $10^{-6}\,{\rm M}_\odot\,{\rm yr}^{-1}$, and the launch speed, $\upsilon\launch$ is usually $\ga100\,{\rm km\,s}^{-1}$. Consequently, convergence requires ${\cal N}\core\ga 10^6$ SPH particles per ${\rm M}_\odot$.

A further issue affecting the runs with continuous outflow is that the high velocity of the launched SPH particles greatly reduces the timestep for these particles and for all those with which they interact, even indirectly -- in particular, those that get entrained in the outflow early on. Even though the code uses individual particle time steps, many particles have short time steps, throughout the simulation, and hence the computing requirements are very demanding. In contrast, the runs with episodic outflow only require these short timesteps for brief periods during and immediately after an outburst. 

%%%%%
\subsection{Varying the physical parameters}\label{chap:ParamStudy}
%%%%%

The Outflow Feedback Model involves six physical parameters, and their values are poorly constrained by observation or theory. In this section we explore how the results change when these parameters are increased or decreased from their default values (i.e. those in the fiducial Run 6, {\sc Episodic:100}: $\theta\wind =0.4\,{\rm rad}$, $r\launch =0.07\,{\rm AU}$, $\theta\jet =0.01\,{\rm rad}$, $f\ejct =0.1$, $r\mn =20\,{\rm AU}$ and $\alpha\MRI =0.1$). In each of Runs 11 through 22 (see Table \ref{Tab:RotIso}), only one physical parameter is changed from its default value, and the numerical parameters are held at their default values, i.e. ${\cal N}\core =10^5$ (hence $m\SPH =10^{-5}\,{\rm M}_\odot$ and $M\res =10^{-3}\,{\rm M}_\odot$) and $\rho\sink =10^{-10}\,{\rm g\,cm}^{-3}$.

%%%%%
\subsubsection{The Wind Opening Angle, $\theta\wind$}\label{chap:theta_max}
%%%%%

We consider three wind opening angles, $\theta\wind =0.6\,{\rm radian}$ (Run 11), $0.4\,{\rm radian}$ (fiducial Run 6) and $0.2\,{\rm radian}$ (Run 12); the evolution of $M\sink(t)$ and $M\tot(t)$ is shown in the top left panel of Fig. \ref{fig:Param_Mass}, and the evolution of $P\outflow(t)$ and $P\tot(t)$ in the top left panel of Fig. \ref{fig:Param_Mom}. If we analyse the results at the end of the simulation, $t\fin$, then, as $\theta\wind$ is decreased from $0.6\,{\rm radian}$ to $0.2\,{\rm radian}$ (i.e. by a factor of 3), there is very little change in $M\sink(t\fin)$ ($\la 1\%$), but the amount of mass escaping, $M\tot(t\fin)$, has increased by $\sim 25\%$; and the amount of momentum escaping, $P\tot(t\fin)$, has increased by $\sim 100\%$. This seemingly counter-intuitive result arises because it is the relatively fast narrow jet, rather than the relatively slow wide-angle wind, that does most of the damage to the core; as $\theta\wind$ is decreased, an increasing fraction of the outflow is concentrated in the jet.

%%%%%
\subsubsection{The Launch Radius, $r\launch$}\label{chap:launching_radius}
%%%%%

We consider three launch radii, $r\launch =0.14\,{\rm AU}$ (Run 13), $0.07\,{\rm AU}$ (fiducial Run 6) and $0.047\,{\rm AU}$ (Run 14; this last value fits the observations of \cite{Lee17}). The evolution of $M\sink(t)$ and $M\tot(t)$ is shown in the top right panel of Fig. \ref{fig:Param_Mass}, and the evolution of $P\outflow(t)$ and $P\tot(t)$ in the top right panel of Fig. \ref{fig:Param_Mom}. Smaller values of $r_{\mathrm{launch}}$ equate to higher outflow velocities (see Eq. \ref{Eq:Vout}), and consequently, all other things being equal, more vigorous feedback. However, as $r\launch$ is decreased, the outburst frequency decreases and the downtime between outbursts lengthens; since outbursts produce abrupt changes in $M\tot$ and $P\tot$, this makes quantitative comparison difficult. If we make the comparison at $\sim 32\,{\rm kyr}$, when all three runs have experienced the same number of outbursts, then, as $r\launch$ is decreased from $0.14\,{\rm AU}$ to $0.047\,{\rm AU}$ (i.e. by a factor of 3), $M\sink$ decreases by $\sim 21\%$, $M\tot$ increases by $\sim 56\%$, and $P\tot$ increases by $\sim 81\%$. By the end of the simulation, at $\sim 50\,{\rm kyr}$, Run 13 (with $r\launch =0.14\,{\rm AU}$) has experienced one more outburst than the other two runs, and the increases in $M\tot$ and $P\tot$ are reduced to $28\%$ and $23\%$ respectively; the decrease in $M\sink$ is still $\sim 21\%$. We conclude that reducing $r\launch$ by a factor of 3 reduces the rate at which $M\sink$ grows by $21\%$, and increases the rate at which mass and momentum escape by $42\pm 14\%$ and $52\pm 29\%$ respectively, with the uncertainty deriving from when the comparison is made and how many outbursts there have been.

%%%%%
\subsubsection{The Jet Opening Angle, $\theta\jet$}\label{chap:theta_0}
%%%%%

\begin{figure*}
\centering
\includegraphics[width=1.0\textwidth]{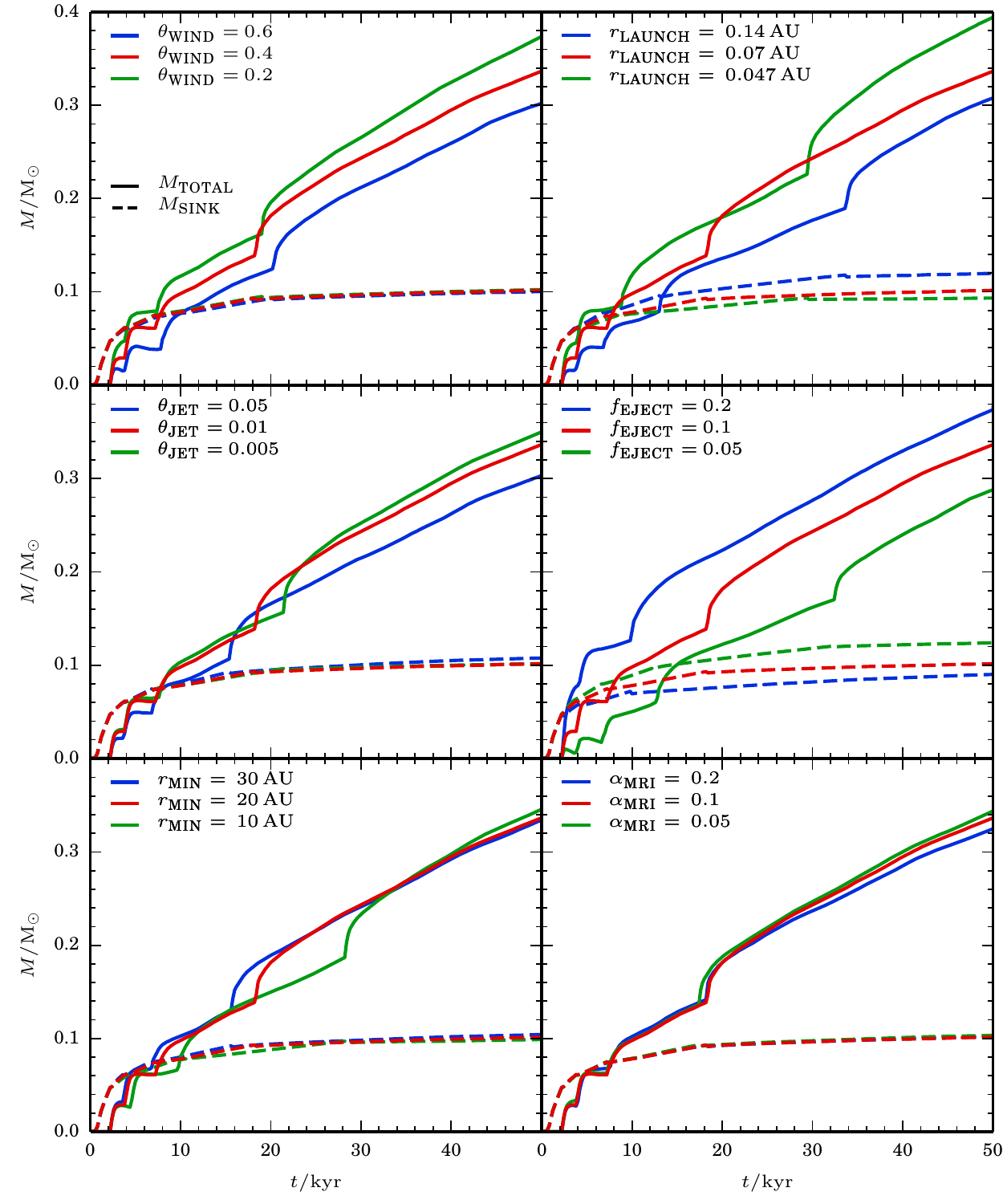}
\caption{The evolution of the sink mass ($M\sink(t)$, dashed lines) and the total mass carried away by all the escaping SPH particles ($M\tot(t)$, solid lines), for different values of the model parameters. {\it Top left:} the wind opening angle, $\theta\wind =0.2\,{\rm radian}$, $0.4\,{\rm radian}$ (default) and $0.6\,{\rm radian}$. {\it Top right:} the launch radius, $r\launch =0.047\,{\rm AU}$, $0.07\,{\rm AU}$ (default) and $0.14\,{\rm AU}$. {\it Middle left:} the jet opening angle, $\theta\jet =0.005\,{\rm radian}$, $0.01\,{\rm radian}$ (default) and $0.05\,{\rm radian}$. {\it Middle right:} the ejected fraction, $f\ejct =0.05$, $0.1$ (default) and $0.2$. {\it Bottom left:} the minimum injection radius, $r\mn =10\,{\rm AU}$, $20\,{\rm AU}$ (default) and $30\,{\rm AU}$. {\it Bottom right:} the Shakura-Sunyayev viscosity parameter, $\alpha\MRI =0.05$, $0.1$ (default) and $0.2$.}
\label{fig:Param_Mass}   
\end{figure*}

\begin{figure*}
\centering
\includegraphics[width=1.0\textwidth]{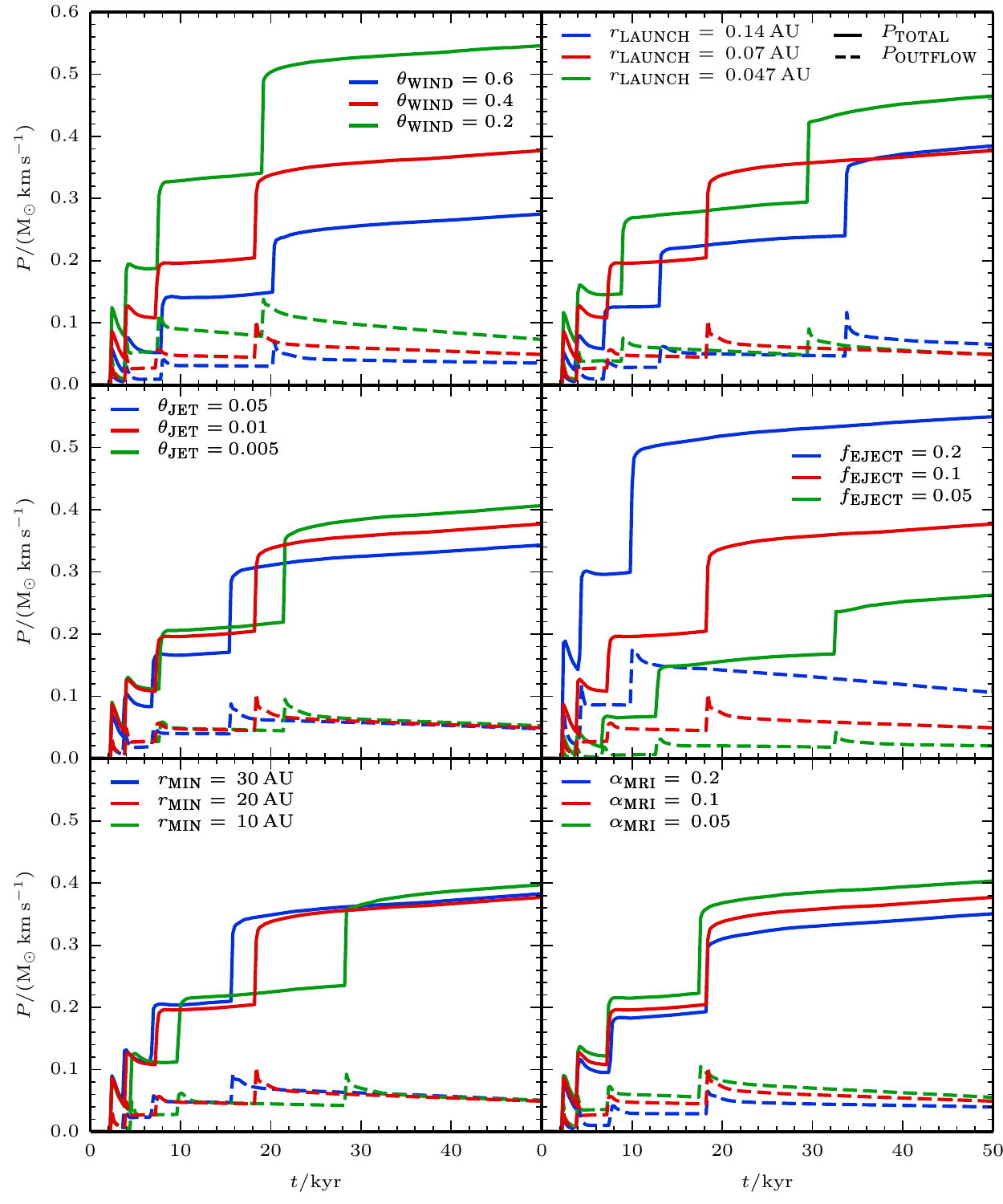}
\caption{As Fig. \ref{fig:Param_Mass} but for the evolution of the momentum carried away by the escaping outflow SPH particles alone ($P\outflow(t)$, dashed lines), and by all the escaping SPH particles (i.e. including ambient SPH particles entrained in the flow; $P\tot(t)$, solid lines).}
\label{fig:Param_Mom}  
\end{figure*}

We consider three jet opening angles, $\theta\jet =0.05\,{\rm radian}$ (Run 15; this is the maximum suggested by \cite{Matzner99}), $0.01\,{\rm radian}$ (fiducial Run 6) and $0.005\,{\rm radian}$ (Run 16). The evolution of $M\sink(t)$ and $M\tot(t)$ is shown in the middle left panel of Fig. \ref{fig:Param_Mass}, and the evolution of $P\outflow(t)$ and $P\tot(t)$ in the middle left panel of Fig. \ref{fig:Param_Mom}. Smaller values of $\theta\jet$ correspond to more tightly collimated, faster jets (see Eqs. \ref{P(mu)} and \ref{Eq:Vout}, and Fig. \ref{fig:VelProfile}), which are slightly more effective in reducing the rate of growth of the sink (lower $M\sink$), and slightly more effective in dispersing the core (larger $M\tot$ and $P\tot$). Specifically, if we make the comparison at the end of the simulations ($\sim 50\,{\rm kyr}$), then, as $\theta\jet$ is reduced from $0.05\,{\rm radian}$ to $0.005\,{\rm radian}$ (i.e. by a factor of 10), $M\sink$ decreases by $\sim 7\%$, $M\tot$ increases by $\sim 15\%$, and $P\tot$ increases by $\sim 19\%$. The timing of outbursts is also not very sensitive to $\theta\jet$, and we conclude that $\theta\jet$ is not a very critical parameter

%%%%%
%%%%%

%%%%%
%%%%%

%%%%%
\subsubsection{The Ejection Fraction, $f\ejct$}\label{chap:ejection_fraction}
%%%%%

We consider three ejection fractions, $f\ejct =0.2$ (Run 17), $0.1$ (fiducial Run 6) and $0.05$ (Run 18). The evolution of $M\sink(t)$ and $M\tot(t)$ is shown in the middle right panel of Fig. \ref{fig:Param_Mass}, and the evolution of $P\outflow(t)$ and $P\tot(t)$ in the middle right panel of Fig. \ref{fig:Param_Mom}. If $f\ejct$ is reduced, the sink grows faster, the amount of mass and momentum escaping from the core goes down, and there are longer downtimes between outbursts. If we make the comparison at the end of the simulations ($\sim 50\,{\rm kyr}$), then, as $f\ejct$ is reduced from $0.2$ to $0.05$ (i.e. by a factor of 4), $M\sink$ increases by $\sim 38\%$, $M\tot$ decreases by $\sim 23\%$, and $P\tot$ decreases by $\sim 52\%$. We conclude that $f\ejct$ is a mildly critical parameter.

%%%%%
\subsubsection{The Minimum Injection Radius, $r\mn$}\label{chap:ejection_radius}
%%%%%

We consider three minimum injection radii, $r\mn =30\,{\rm AU}$ (Run 19), $20\,{\rm AU}$ (fiducial Run 6) and $10\,{\rm AU}$ (Run 20). The evolution of $M\sink(t)$ and $M\tot(t)$ is shown in the bottom left panel of Fig. \ref{fig:Param_Mass}, and the evolution of $P\outflow(t)$ and $P\tot(t)$ in the bottom left panel of Fig. \ref{fig:Param_Mom}. Reducing $r\mn$ from $30\,{\rm AU}$ to $10\,{\rm AU}$ (i.e. by a factor of 3) has very little effect, apart from increasing the downtime between outbursts; at the end of the simulations ($\sim 50\,{\rm kyr}$) $M\sink$, $M\tot$ and $P\tot$ all differ by at most $\sim 5\%$. We conclude that $r\mn$ is not a critical parameter.

%%%%%
\subsubsection{The Shakura-Sunyayev Viscosity Parameter, $\alpha\MRI$}\label{chap:alpha}
%%%%%

We consider three Shakura-Sunyayev viscosity parameters, $\alpha\MRI =0.2$ (Run 21), $0.1$ (fiducial Run 6) and 0.05 (Run 22). The evolution of $M\sink(t)$ and $M\tot(t)$ is shown in the bottom right panel of Fig. \ref{fig:Param_Mass}, and the evolution of $P\outflow(t)$ and $P\tot(t)$ in the bottom right panel of Fig. \ref{fig:Param_Mom}. In the \cite{Stamatellos12a} prescription for episodic accretion, $\alpha\MRI$ controls the accretion rate from the IAD onto the protostar during an outburst. Reducing $\alpha\MRI$ leads to longer, but less intense, outbursts, and the net effect is rather small. If we make the comparison at the end of the simulations ($\sim 50\,{\rm kyr}$), then, as $\alpha\MRI$ is reduced from $0.2$ to $0.05$ (i.e. by a factor of 4), $M\sink$ increases by $<2\%$, $M\tot$ by $<6\%$, and $P\tot$ by $<16\%$. We conclude that $\alpha\MRI$ is not a critical parameter.

%%%%%
\subsubsection{Self-regulated outflow feedback}\label{chap:selfReg}
%%%%%

The most critical physical parameters appear to be $\theta\wind$ (because it influences the extent to which the outflow is concentrated in the jet), $r\launch$ (because it influences the velocity at which the outflow is launched) and $f\ejct$ (because it influences the amount of mass going into the outflow). Furthermore, the quantity that is most sensitive to these parameters is $P\tot$, and $M\sink$ is least sensitive. However, all the dependences are `sub-linear'. For example, the three most extreme dependences are 
\begin{eqnarray}
&\frac{d\ln\left(P\tot\right)}{d\ln\left(\theta\wind\right)}&\;\;\,\sim\;\;-\,0.63\,,\\
-0.54\;\,\la\;\,&\frac{d\ln\left(P\tot\right)}{d\ln\left(r\launch\right)}&\;\,\la\;\, -0.19\,,\\
-0.40\;\,\la\;\,&\frac{d\ln\left(M\tot\right)}{d\ln\left(r\launch\right)}&\;\,\la\;\,-0.22\,.
\end{eqnarray}
The basic reason for these rather weak dependences is a self-regulation mechanism. In runs with parameters that reduce the impact of the outflow on the core, the sink is able to accrete faster, leading to more frequent outbursts, and shorter downtimes between outbursts. We conclude that using the default parameter values will not significantly affect our main findings.

%%%%%
\section{The Turbulent Setup}\label{chap:Results}
%%%%%

%%%%%
\begin{figure*}
\centering
\includegraphics[width=1.0\textwidth]{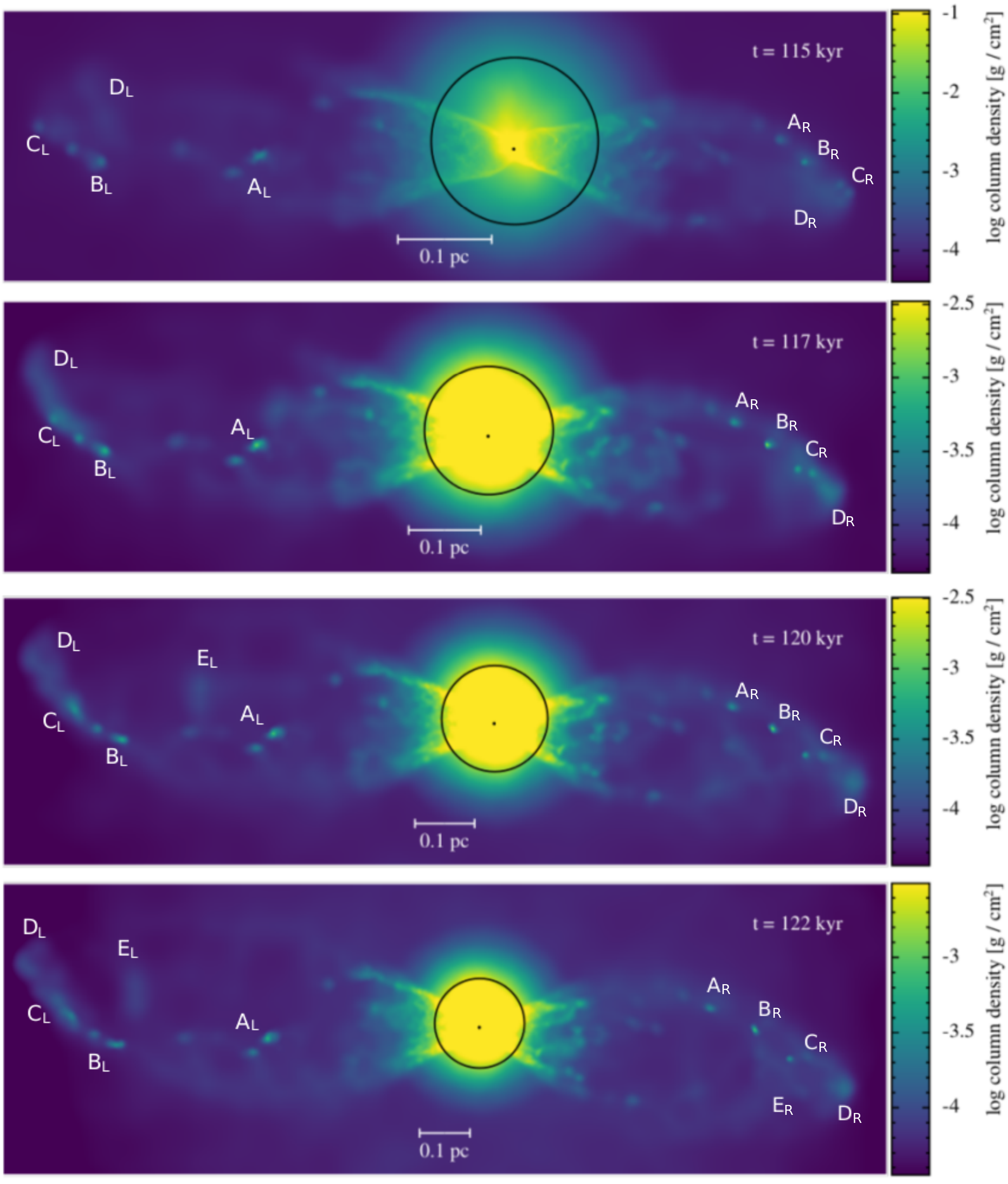}
\caption{Column density images of the Turbulent Setup, at $t=115,\;117,\;120\;{\rm and}\;122\,{\rm kyr}$.  The circle indicates the initial radius of the core, and the viewing angle is chosen so that the outflow axis is approximately horizontal. The bullets are labelled A$\rmsc{L/R}$ through E$\rmsc{L/R}$ for the corresponding left and right lobe, in order of increasing age (see Fig. \ref{Fig:BES_accRate}). The time series, from top to bottom, shows the end of a cycle of a bullet, here bullet D$\rmsc{L}$, when it hits the leading shock front. Bullet D$\rmsc{L}$ hits the leading shock front with high velocity, overtakes the older bullets and decelerates. In the last panel, the newly ejected bullet E$\rmsc{L}$ is in a comparable position as former bullet D$\rmsc{L}$ in the first frame.}
\label{Fig:BES_606}
\end{figure*}
%%%%%

%%%%%
\begin{figure*}
\centering
\includegraphics[width=1.0\textwidth]{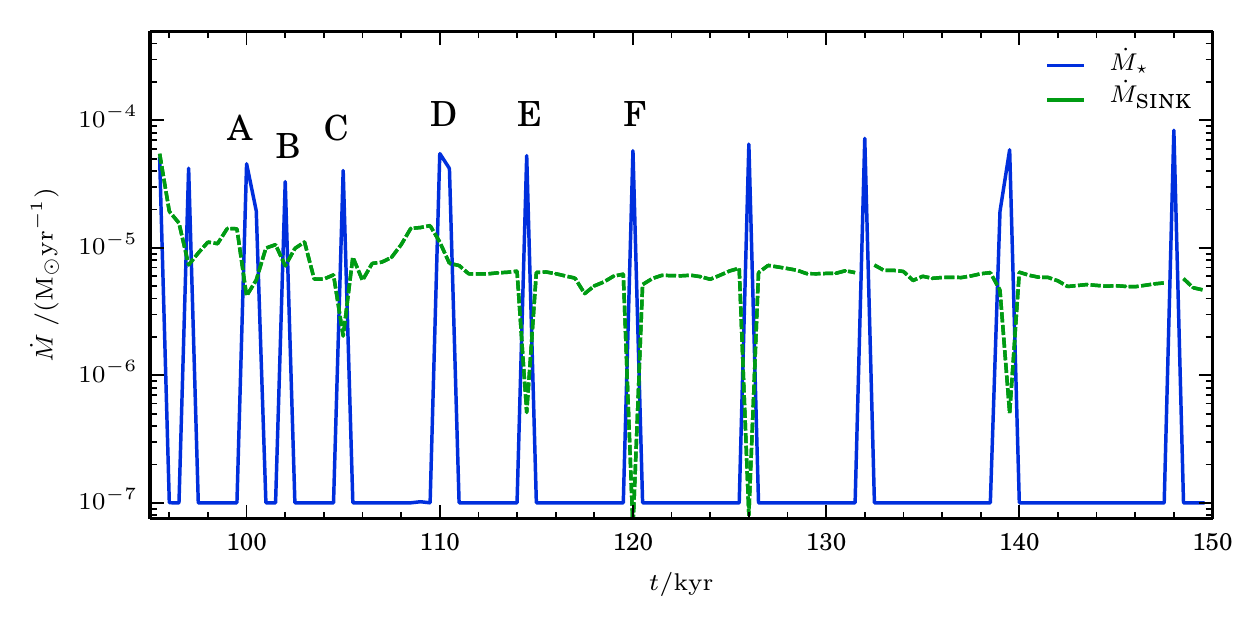}
\caption{Evolution of the accretion rates onto the sink particle, $\dot{M}\sink(t)$ (green), and the protostar, $\dot{M}\astr(t)$ (blue). The labels, A to F, correspond to the outflow bullets formed in the individual outbursts (see Fig. \ref{Fig:BES_606} and \ref{fig:MV-PV-Relation}). Note that the time resolution in this figure is too coarse to fully resolve the very short outbursts.}
\label{Fig:BES_accRate}
\end{figure*}
%%%%%

In order to illustrate a more realistic situation, Fig. \ref{Fig:BES_606} shows column-density images from a run with the Turbulent Setup (see Section \ref{SEC:setup}). All parameters for the Episodic Accretion Model (Section \ref{chap:episodic_acc}) and the Outflow Feedback Model (Section \ref{SEC:Outflow}) are the same as in the Rotating Setup fiducial Run 6 (see Table \ref{Tab:RotIso}). A single protostar forms at $\sim 90\,\rm{kyr}$ with a final mass of $M\astr =0.45\,\rm{M}_\odot$. Its accretion ceases at $\sim 180\,\rm{kyr}$ since by this stage most of the gas has either been accreted or dispersed. The simulation is finally stopped at $\sim 230\,\rm{kyr}$. 

The protostar launches outflow bullets labelled A$\rmsc{L/R}$ through E$\rmsc{L/R}$ corresponding to the left and right outflow lobe. In the remainder of the paper, we will only refer to the left lobe and therefore we drop the subscript L henceforth. The outflow bullets form an S-shaped chain of Herbig--Haro objects. The S-shape is due to the varying orientation of the angular momentum, $\boldsymbol{L}\IAD$, caused by anisotropic accretion onto the IAD from core \citep{Ybarra06, Wu09, Zhang13, Frank14}. The first two bullets interact with the collapsing dense core material and are thus not visible in Fig. \ref{Fig:BES_606}. The timespan from 115 kyr to 122 kyr is chosen to capture the end of a cycle of an outflow bullet, here bullet D, when it hits the leading shock front. This cycle happens in a similar fashion for all other bullets, except the first bullet A. This is the first bullet that breaks out of the core and survives as a coherent structure. Bullets B and C have higher velocity and have overtaken A. Bullet D starts off faster still, and by the final frame it has overtaken all the others and hit the leading shock, where it is slowed down. Bullet E is even faster, but by the final frame it has not yet caught up with D and is in a similar position as former bullet D in the first frame. A sixth bullet has been launched, but is still inside the core, and therefore can not be seen on Fig. \ref{Fig:BES_606}. 

Fig. \ref{Fig:BES_accRate} shows the accretion rates onto the sink, $\dot{M}\sink$ and onto the protostar, $\dot{M}\astr$, between $t=90\,\rm{kyr}$ (when the protostar forms) and $t=150\,\rm{kyr}$. $\dot{M}\sink$ is relatively constant, between $\sim 2\times 10^{-5}$ and  $\sim 5\times 10^{-6} \,\rm{M_\odot\,yr}^{-1}$. The accretion rate tends to decrees over time.
It drops briefly, following an outburst, because the outburst heats the accretion disc \citep[cf.][]{Lomax14, Lomax15}. The accretion rate of the protostar is low in the quiescent phase, $\dot{M}\astr =10^{-7}\,\rm{M_\odot\,yr}^{-1}$, but approaches $10^{-3}\,\rm{M_\odot\,yr}^{-1}$ during an outburst. Note that the time resolution in Figure \ref{Fig:BES_accRate} is too coarse to resolve the very short outbursts. The downtime between outbursts is $\sim 4000\pm 2000\,\rm{yr}$ and the duration of an outburst is $\sim 40\pm 20\,\rm{yr}$ \citep{Stamatellos07, Stamatellos11}.

%%%%%
\begin{figure*}
\centering
\includegraphics[width=1.0\textwidth]{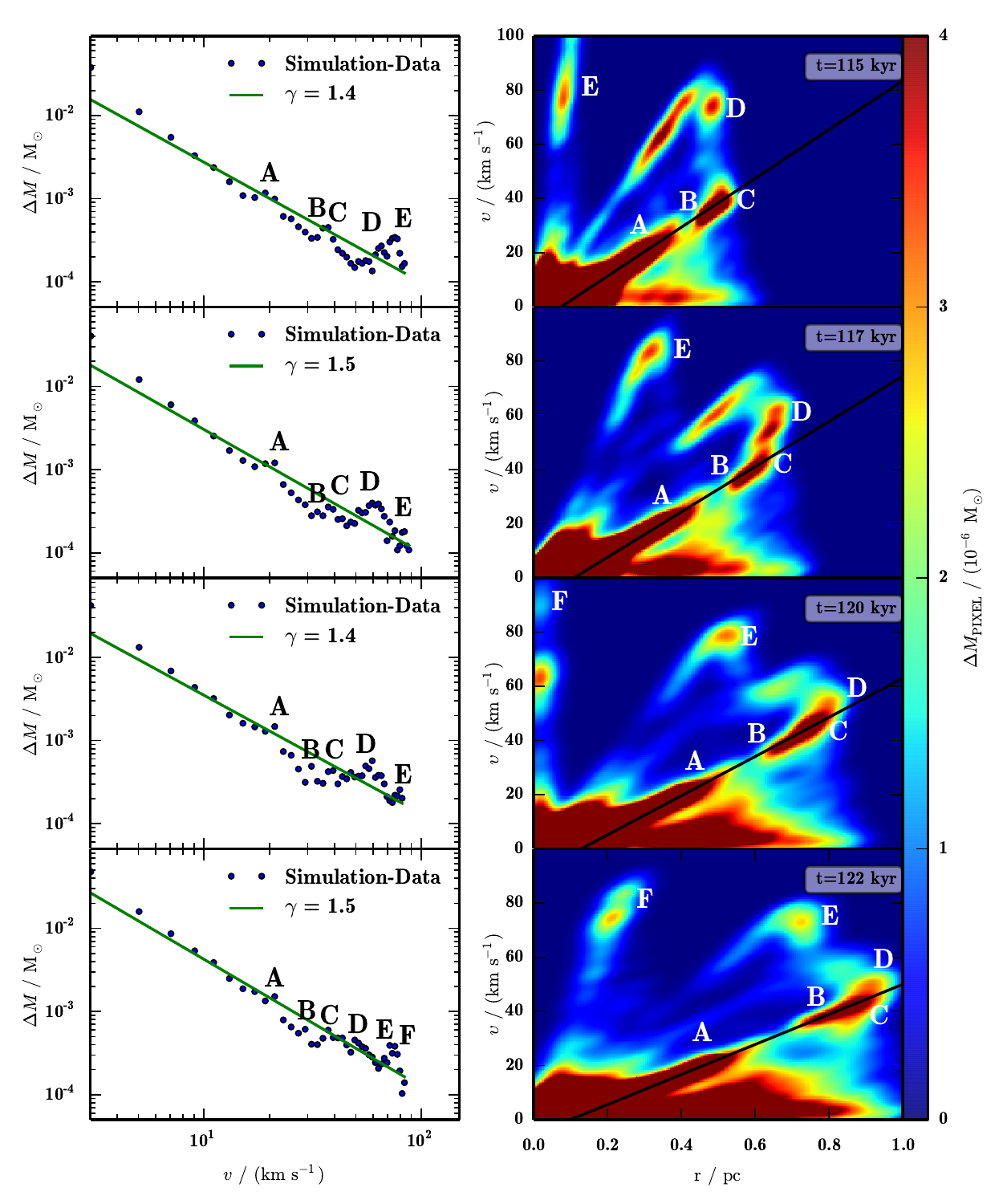}
\caption{ MV- and PV-diagrams for the lefthand outflow lobe, at the same four times illustrated in Fig. \ref{Fig:BES_606}, and using the same labels (A through F) to identify features associated with the individual bullets. {\it Left Panels:} MV-diagrams showing power-law behaviour with slope $\gamma\sim 1.5$. The locations of individual bullets are marked by small jumps on this plot. {\it Right panels:} PV-diagram. The black line shows the best-fit Hubble law ($\upsilon\rad\sim\upsilon\subO+\upsilon'\subO r$), and marks the location of bullets that have already been decelerated at the leading shock; those bullets that have not yet been decelerated lie above this line.}
\label{fig:MV-PV-Relation}
\end{figure*}
%%%%%

%%%%%
\subsection{Position-velocity diagrams}
%%%%% 

In order to determine whether the simulated outflow reproduces the `Hubble-Law' relation, and individual `Hubble Wedges', as observed, for example, by \cite{Bachiller90, Arce01, Tafalla04, Garcia09, Wang14}, we construct PV-diagrams, i.e. false-colour plots of
\begin{eqnarray}
\Delta M_{_{\rm PIXEL}}& = &\frac{d^2\!M}{dr\,d\upsilon\rad}\;\Delta r\;\Delta\upsilon\rad\,
\end{eqnarray}
on the $(r,\upsilon\rad)$ plane. Here, $r$ is radial distance, and $\upsilon\rad$ is radial velocity, both measured from the centre of mass. For the velocity axis we take 180 bins with a width of $ \Delta\upsilon\rad = 0.56\,\mrm{km} \, \mrm{s}^{-1}$. For the spatial axis, we take 180 spherically symmetric shells around the centre of mass with a width of $\Delta r = 0.0056\,{\rm pc}$ and consider only particles with radial outflow velocities larger than the escape velocity (Eq. \ref{Eq:OutfowCrit}). Note that in this case $M\tot = 3.9 \, {\rm M}_\odot$, so the escape velocity at $0.1\,{\rm pc}$ is $0.58\,\rm{km\,s}^{-1}$. This effectively constrains our analysis to the outflow cones. As a final step, we use a kernel density estimator to compute $\Delta M_{_{\rm PIXEL}}$, with a smoothing length obtained using "Scott's Rule" \citep{Scott92}. 

Fig. \ref{fig:MV-PV-Relation} (right hand side) shows PV diagrams for the left lobe of the outflow at the same times as shown on Fig. \ref{Fig:BES_606}. Features associated with the individual bullets are marked A, B, C, etc., in both figures.

The PV-diagram at $115\,\rm{kyr}$ is shown in the top right panel of Fig. \ref{fig:MV-PV-Relation}. At this stage, bullets A, B and C define the Hubble Law. The oldest bullet, A, moves slowest, at $\sim 25\,\rm{km\,s}^{-1}$ and has only reached $\sim 0.3\,{\rm pc}$. Bullets B an C appear merged, but are in fact kinematically separate. B moves outwards faster than A, at  $\sim 30\,\rm{km \,s}^{-1}$, and has reached $\sim 0.46\,\rm{pc}$. C moves outwards faster than B, at  $\sim 40\,\rm{km\,s}^{-1}$, and has reached $\sim 0.52\,\rm{pc}$. Bullet D moves outwards at $\sim 80\,\rm{km\,s}^{-1}$, but has only been going for $\sim 6\,\rm{kyr}$, so it has not yet hit the leading shock and been decelerated. Bullet E has only just been launched, within the last $\sim 1\,\rm{kyr}$; the outburst at $\sim 114\,\rm{kyr}$ that launches E can be seen on Fig. \ref{Fig:BES_accRate}.

At subsequent times (reading down the righthand column of Fig. \ref{fig:MV-PV-Relation}: $117\,\rm{kyr}$, $120\,\rm{kyr}$ and $122\,\rm{kyr}$) we see Bullet D decelerate and line up with the Hubble Law; Bullet E start to hit the leading shock and decelerate, and the launch of Bullet F; the outburst at $\sim 120\,\rm{kyr}$ that launches F can be seen on Fig. \ref{Fig:BES_accRate}. These results are very similar to the $350$ GHz continuum and CO $J=3\!-\!2$ observations of the outflow from IRAS 04166+2706 reported by \cite{Wang14}. They find numerous high velocity outflow bullets, which slowly decelerate as they move outwards.

%%%%%
\subsection{Mass-Velocity-Relation}\label{chap:MV}
%%%%%

The lefthand panels of Fig. \ref{fig:MV-PV-Relation} show MV-diagrams for the left lobe of the simulated outflow, i.e. plots of 
\begin{eqnarray}
\Delta M&=&\frac{dM}{d\upsilon\rad}\;\Delta\upsilon\rad
\end{eqnarray}
against $\upsilon\rad$, at the same times as the PV-diagrams in the righthand panels. Each point represents the mass in a radial velocity interval $\Delta\upsilon\rad =2\,\rm{km\,s}^{-1}$, and only points that correspond to $\geq 22$ SPH articles ($\geq 10^{-4}\,\rm{M}_\odot$) are considered. At low velocities, $\upsilon\rad\la\,80\,\rm{km\,s}^{-1}$, the plots can be fitted well with a single shallow power-law, 
\begin{eqnarray}
\frac{dM}{d\upsilon\rad}&\propto&\upsilon\rad^{-\gamma}\,,
\end{eqnarray}
with $\gamma\simeq 1.5$, in good agreement with observed and simulated values \citep{Lada96, Kuiper81, Bachiller99, Richer00, Keegan05,  Arce07, Liu17, Li18}.

Some observers have reported a knee, at high velocities, above which the slope abruptly becomes much steeper. We find no evidence for this, possibly because our simulation is unable to resolve the small amounts of mass involved.

On the MV diagram, the bullets are manifest as small local peaks. The oldest bullet, A, has the lowest velocity, and the youngest bullet, F, has the highest. There are two reasons for this. (a) The protostellar mass, $M\astr(t)$ increases, and so the later bullets are launched at higher velocity (see Eq. \ref{Eq:Vout}). (b) The later bullets encounter less resistance because earlier bullets have cleared the way for them.

%%%%%
\section{Discussion}\label{chap:Discussion}
%%%%%

Using our new model, we are able to simulate the collapse of a core to form a protostar, and the role of episodic accretion and outflow in regulating the growth of the protostar and launching high-velocity bullets into the surroundings. This produces kinematic features that mimic the Hubble Law and Hubble Wedge features seen in real star-forming cores with outflows, and allows us to connect these features in an evolutionary sequence. The bullets launched later have higher velocities, up to $\sim 120\,\rm{km\,s}^{-1}$; this is basically because the mass of the protostar increases, and hence the escape speed from its locality increases. They are also launched into a cavity that has been cleared out by earlier bullets \citep{Wang14}, and therefore they tend to travel further before they run into the ambient medium and are decelerated. As a result the later bullets leave wakes pointing back towards the star -- these are the Hubble Wedges -- and the earlier bullets define an approximately linear Hubble-like velocity field. In our simulations, the individual outflow events also produce bumps in the MV relation that are similar to those observed by \cite{Qiu09}

The main difference between the prescriptions for ejection used by e.g. \cite{Federrath14, Offner14, Offner17} and our model is that in our model the ejection rate is not determined directly by the accretion rate onto the sink, but by the accretion rate onto the central protostar within the sink, and this rate is moderated by an episodic cycle. Nonetheless, like \cite{Federrath14} we find that the feedback is self-regulated, in the sense that the outflow properties do not depend strongly on the ejection fraction $f\ejct$.

By varying $\theta\jet$ we find, like \cite{Offner14}, that even highly collimated jets are able to entrain large fractions of gas. In contrast to their turbulent simulation, we find a significant precession of the outflow, leading to an S-shaped chain of Herbig--Haro objects. Compared with \cite{Federrath14} and \cite{Offner17} we find a much higher ratio of entrained to ejected gas, $\sim 30$, because we eject gas at higher velocities of up to $120\,\rm{km\,s}^{-1}$. This also explains our lower star formation efficiency, of only $\sim 10\%$.

We note the following shortcomings of our Outflow Feedback Model. Unlike \cite{Offner17}, we do not use a stellar evolution model, and therefore we have to invoke a somewhat more arbitrary prescription for the outflow velocity, and for the amount of angular momentum that is carried away by the outflow; specifically, we assume a constant fraction of the angular momentum is removed \citep{Herbst07,Bouvier14}. A stellar evolution model would also improve the treatment of radiative feedback from the protostar -- in the sense that the current treatment probably underestimates the protostellar luminosity between outbursts -- and we are planning to include the protostellar model of \cite{Offner17} in a follow-up paper \citep{RohdePrep}. 

In addition, although the treatment of radiation transport \citep{Stamatellos07, Stamatellos11} appears to  work reasonably well in the context of collapsing cores, and even accretion disks, it is unlikely to work so well in the walls of the outflow cavity \citep{Kuiper16}. However, since we are simulating the formation of low mass stars, the effects of radiation are expected to be small compared with the outflow feedback.

Finally, the role of the magnetic field is implicit in the model for accretion onto the protostar and the outflow launching prescription, but its effect on the dynamics of core collapse and the interaction between the core and the outflow is ignored here, and could be significant \citep{Commercon10,Seifried12, Wurster16, Wurster18}.

In future work we will use the model presented in this paper to study the influence of episodic outflow feedback on the star formation efficiency and the shape of the stellar initial mass function. With higher resolution, we will be able to study the high-velocity tail of the PV-diagram, which is not resolved in the simulations presented in this study. These high-resolution simulations will also enable us to examine the rotational properties of outflow bullets, and to compare them with the observations of e.g. \cite{Launhardt09, Chen16, Lee17} and \cite{Tabone17}. 
%%%%%
\section{Conclusion}\label{chap:Conclusion}
%%%%%

Recent studies of outflows from young, low-mass protostars suggest that their accretion and outflows are episodic. Our newly developed sub-grid outflow model for SPH takes this episodic behaviour into account. Besides the benefits of a more realistic outflow model, episodicity actually  decreases the mass-resolution needed to properly model the outflow (because the gas is denser in an outflow burst than in a continuous outflow), and at the same time reduces the computation required (because the outflow particles are launched in short outbursts).

We explore the effect of episodic accretion and outflow from a protostar formed at the centre of an initially static, rigidly rotating, spherically symmetric core with $\rho\propto r^{-2}$. We show that key properties like the rate of growth of the protostar, and the net mass and momentum carried away by the outflow, are only weakly dependent on numerical parameters. In particular, reliable results can probably be obtained with quite low mass-resolution ($m\SPH\sim 2\times 10^{-5}\,\rm{M}_\odot$) and low sink creation density ($\rho\sink\sim 10^{-12}\,\rm{g\,cm}^{-3}$). The rate of growth of the protostar, and the net mass and momentum carried away by the outflow, are also rather weakly dependent on the physical parameters, because of self-regulation: if the physical parameters are changed so as to increase the outflow driven by a given rate of accretion onto the protostar, then the rate of accretion is reduced by the outflow, so the actual rate of outflow is little changed. We conclude that our model can be implemented in large-scale simulations of molecular clouds, where $m\SPH$ has to be large, and $\rho\sink$ has to be small, in order to follow simultaneously the formation of many protostars, all potentially having outflows.
	
We follow the effect of episodic accretion and outflow from a protostar formed near the centre of an initially turbulent, spherically symmetric core with density profile proportional to that of a Bonnor-Ebert sphere. The episodic outflow produces a parsec-long S-shaped chain of bullets, which we identify with Herbig-Haro objects. The position-velocity diagram for these bullets shows two features. Bullets that were ejected early were ejected with lower radial velocities, and have by now been decelerated at the leading shock; these bullets have radial velocities that are linearly proportional to their radial distance, i.e. they subscribe to a Hubble Law. Bullets that were only ejected recently have higher radial velocities and have not yet been decelerated at the leading shock front; these bullets form Hubble wedges. The mass-velocity relation for the gas in the outflow can be fit approximately with $dM/d\upsilon\rad\propto\upsilon\rad^{-1.5}$, in good agreement with observation. Individual bullets are manifest as small bumps along this relation. If there is a steeper slope at higher velocities, we are unable to resolve it in this simulation.

\section*{acknowledgements}
The authors like to thank the anonymous referee for the comments that helped to significantly improve the paper. 
PR, SW and SC acknowledge support via the ERC starting grant No.
679852 'RADFEEDBACK'. 
PR, SW, and DS acknowledge the support of the Bonn-Cologne Graduate School,
which is funded through the German Excellence Initiative. 
DS, and SW thank the DFG for funding
via the SFB 956 'Conditions \& impact of SF'.
APW gratefully acknowledges the support of a consolidated
grant (ST/K00926/1) from the UK Science and Technology Facilities Council. 
The authors gratefully acknowledge the Gauss Centre for Supercomputing e.V. 
(www.gauss-centre.eu) for funding this project by providing computing time on the GCS Supercomputer SuperMUC at Leibniz Supercomputing Centre (www.lrz.de).
PR acknowledges D. Price for providing the visualization tool SPLASH \citep{Price11}.

%\renewcommand{\leftmark}{\sc References}
%\bibliography{references}
\bibliographystyle{mnras}
\bibliography{references}
%\clearpage
% Don't change these lines

%%%%%%%%%%%%%%%%%%%%%%%%%%%%%%%%%%%%%%%%%%
%%%%%%%%%%%%%%%%%%%%%%%%%%%%%%%%%%%%%%%%%%

\bsp	% typesetting comment
\label{lastpage}
\end{document}